\documentclass [twocolumn] {svjour3}

\usepackage{graphicx}
\usepackage{subfigure}
\usepackage{bm}
\usepackage{amssymb}
\usepackage[latin1]{inputenc}
\usepackage[portuges,frenchb,english]{babel}
\usepackage{color}
\usepackage[ruled,vlined]{algorithm2e}

\def\eg{{\em e.g., }}
\def\ie{{\em i.e., }}

\newcommand{\revisionter}[1]{{#1}}
\def\be{\begin{equation}}
\def\ee{\end{equation}}
\def\ba{\begin{array}}
\def\ea{\end{array}}
\def\ban{\begin{eqnarray}}
\def\ean{\end{eqnarray}}

\journalname{J. Math. Imag. Vision, (JMIV-S-12-00190)}

\begin{document}

\title{Efficient binary tomographic reconstruction}

\author{St\'{e}phane Roux \and Hugo Leclerc \and Fran\c{c}ois Hild}

\institute{St\'{e}phane Roux \and Hugo Leclerc \and Fran\c{c}ois Hild \at LMT-Cachan
(ENS de Cachan/CNRS/UPMC/PRES UniverSud Paris)
        61 avenue du Pr\'esident Wilson, F-94235 CACHAN (FRANCE)\\
        \email{\{stephane.roux,hugo.leclerc,francois.hild\}@lmt.ens-cachan.fr}}

\date{Received: date / Accepted: date}

\maketitle

\begin{abstract}
Tomographic reconstruction of a binary image from few projections
is considered. A novel {\em heuristic} algorithm is proposed, the
central element of which is a nonlinear transformation
$\psi(p)=\log(p/(1-p))$ of the probability $p$ that a pixel of the
sought image be 1-valued. It consists of backprojections based on
$\psi(p)$ and iterative corrections. Application of this algorithm
to a series of artificial test cases leads to exact binary
reconstructions, (\ie recovery of the binary image for each single
pixel) from the knowledge of projection data over a few
directions. Images up to $10^6$ pixels are reconstructed in a few
seconds. A series of test cases is performed for comparison with
previous methods, showing a better efficiency and reduced
computation times.
\end{abstract}


\section{Introduction}
\label{se:intro}

Tomographic reconstruction is a mature
topic~\cite{Kak01,Gardner06} for which a variety of algorithms is
now available from the celebrated Filtered BackProjection
(FBP)~\cite{FBP}, including its fast
implementation~\cite{Basu01a,Basu01b}, to various Algebraic
Reconstruction Techniques (ART)~\cite{Herman76}. Numerous
extensions have been proposed for enhanced accuracy or speed.  ART
appears as the method of choice for a small number of projections
or noisy data, whereas FBP is superior in terms of computation
times.

A specific class of reconstruction problems appears when the image
to be reconstructed has only few gray levels (termed discrete
reconstruction) or is binary (\ie its ``pixels'' are either black
or white)~\cite{Herman99,Herman07}. This additional piece of
information is extremely valuable and allows dealing with very few
projections, provided the image to be reconstructed is
sufficiently ``simple.''

However, the discrete nature of the image to be reconstructed
renders more difficult the recourse to classical reconstruction
algorithms. In order to reach acceptable reconstructions (perfect
ones are generally out of reach), optimization techniques specific
to NP-complete problems (binary reconstruction has been shown to
belong to this class of ``hard'' problems~\cite{NPComplete}) were
considered, namely, Simulated Annealing (SA)~\cite{Weber06,Liao07}
or Genetic Optimization (GO)~\cite{Varga10} have been proposed and
shown to be able to capture approximately synthetic phantoms over
images of size $N\times N$, with $N \le 256$~pixels, and a number
$M$ of projections in the range from 4 to 10, within a computation
time of order of a few hours (even for recent GPU
implementations~\cite{Varga10}). Smaller $N$ values, say of order
64~pixels, still require a computation time of the order of a few
minutes in those references. Such performances rely on the fact
that the image contains few connected domains generally with
smooth boundaries. One additional difficulty of such methods is
their rather slow convergence and sensitivity to the way the
algorithm is driven. In addition, there is no guarantee that the
algorithm is not trapped in a local minimum.

In such a context, an iterative correction approach seems more
appealing, and recent works have led to much more successful
results either in terms of quality of the reconstruction and
computation  cost~\cite{Batenburg07,Batenburg08}. The binary
nature of the sought image, although offering severe constraints
that are helpful to compensate for the lack of information,
suggests that combinatorial type approaches are needed.  Thus, the
challenge is to conciliate corrections, which naturally use a
continuous representation of the image, and the known {\it a
priori} information on its binary character. Early
attempts~\cite{Censor} to interleave classical reconstruction
steps, and ``binary steering'' favoring 0 or 1 for pixel values
showed encouraging results, yet $64\times 64$ images could not be
exactly reconstructed after 1000 iterations.

The binary nature of the image constitutes by itself
an element of simplicity, namely, a single bit of information is
needed per pixel.  However, additional elements can further
simplify the problem of reconstructing the image from a few
projections.  In particular, the notion of ``sparsity'' has been
shown in the recent years to allow for very efficient restoration
of signals (or here images) from few partial measurements or data.
Yet, sparsity may have different appearance.  At least two
categories can be distinguished:\\
- {\it Type A}. If no spatial correlation exists in the image,
then sparsity has to rely only on the number of 1-valued pixels.
It has been shown in a series of work starting from
Ref.~\cite{Donoho04} on undersampling theory, that a
``phase-transition'' exists separating solvable problems from
unsolvable ones depending on the quantity of available
information.  In the framework of sparsity based on the
\revisionter{density} of non-zero pixels only, the order of
magnitude of the maximum concentration of non zero pixels
\revisionter{(\ie number of pixels relative to the image size)},
$p_c$, in a $N\times N$ pixel image, based on $M$ projections has
been shown by Donoho and Tanner~\cite{Donoho10a,Donoho10b} to
amount to
    \be\label{eq:pc_Donoho}
        p_c \approx \frac{M}{2N \log(N/M)}
    \ee
\revisionter{It may be convenient to introduce the undersampling
ratio, $\delta\equiv N/M$, so that $p_c\approx
[2\delta\log(\delta)]^{-1}$.} To mention an example, for a
1-Mpixel image, $N=1000$, and $M=10$ projections, $p_c$ is of the
order of $10^{-3}$, or about one non-zero pixel (at most) per
line. This threshold is a theoretical level, yet available
algorithms~\cite{Jafarpour} allow such problems to be addressed
with an effective threshold that is quite close to the theoretical
value.  \revisionter{Note that for $p>p_c$ still a unique binary
solution may exist, but there is no guarantee that this solution
can be reached through a convex minimization procedure.}\\
- {\it Type B}.  Spatial correlations may also contribute to the
reduction of ``complexity'' in the image.  If all 1-valued pixels
are grouped into a few blobs, the information content of the image
is reduced and hence an arbitrary proportion of black or white
pixels can be dealt with, and still a small number of projections
is needed to reconstruct the image. Along this line, Cand\'{e}s
{\it et al.}~\cite{CS_Tomo} introduced an algorithm to reconstruct
an image from a set of projections provided the image consists of
a few number of domains, each of which having a uniform gray
level. The key to the solution is here again a low level of
``complexity'' in the image. However, in contrast with the
previous case, spatial correlations are here essential. In this
case, the image gradient is a sparse field, and a regularization
based on the ``total variation'' is essential to reach the
solution. \revisionter{Although recent works~\cite{Needell13}
progress toward solid results on stable and robust image recovery
from noisy data based on total variation minimization, yet a proof
of a  phase transition for type B problems (comparable to that of
type A) is not currently available.}

In the following, all examples are believed to belong
to the second class of problems.  However, no practical definition
of the relevant measure of ``complexity'' can be formally derived
as above mentioned for Type A. Based on the above result, an
attempt to transpose this measure for the studied cases is
provided in Sect.~\ref{Sec:Complexity}.  Let us also note that
Gouillart {\it et al.}~\cite{Gouillart} have recently proposed a
different type of algorithm based on ``message passing'' for
binary and discrete tomography reconstruction where it was
suggested that the proportion of 0-1 pixel pairs (or interfaces)
is responsible for the image ``complexity''. The proposed
algorithm is essentially heuristic, and no guarantee for
convergence is proposed.  When too few projections are given, the
algorithm does not converge and the residual difference with the
sought image fluctuates around a constant value which may not be
small. However, when applied to test images which have been
proposed earlier~\cite{Batenburg07}, the reconstruction
performance appears as superior both in terms of quality and
time.

In terms of applications, many different fields are concerned.
Medical imaging is one of the most demanding
applications~\cite{Medical,Medical1}, and here the interest lies
in the X-ray dose reduction for the patient, provided the
reconstruction can be limited to two phases.  In the field of
fluid mechanics, Tomographic Particle Image Velocimetry (Tomo-PIV)
reconstruction of tracer particle distribution in a suspending
transparent fluid is required from optical images taken by few
cameras~\cite{PIV1,PIV2}, hence few projections.
Finally, in the field of materials science, an amazing progress
has been achieved in the recent years in terms of fast data
acquisition~\cite{FastTomo}. Full 3D scans can now be acquired in
less than one second in large scale synchrotron facilities.
Whenever the microstructure could tolerate a simple representation
as a binary image, the reduction in the number of projections
could automatically be translated into a larger scanning rate, or
finer temporal resolution.  These are three examples where
progress in the reconstruction using reduced projections would be
very rewarding.

Section~\ref{Sec:Pb} defines the problem of tomographic
reconstruction and presents some regularization strategies
classically used in this context. Section~\ref{Sec:Algo}
introduces the algorithm used, and a first illustration is shown
in Section~\ref{sec-illus}. Section~\ref{Sec:Multiscale}
introduces a multi-scale formulation that enhances the efficiency
of the algorithm. A detailed comparison with some benchmark tests
is reported in Section~\ref{Sec:TestCases}. An attempt to
rationalize the performances of the text as a function of a
proposed measure of complexity is proposed in
Sect.~\ref{Sec:Complexity}, and the effect of noise on
reconstruction is documented for a single image in
Sect.~\ref{Sec:Noise}. Finally, Section~\ref{Sec:Conclusion}
recalls the major results and some perspectives are discussed.

\section{Statement of the problem}\label{Sec:Pb}

The problem consists of identifying a binary-valued discrete image
$f(\bm x)$ where $f$ is valued in $\{0,1\}$.  $f$ is defined at
pixels $\bm x$ having integer coordinates, inside a  2D domain,
$\bm x\in {\cal D}$, chosen in the sequel to be a circular disk of
radius $N/2$ where $N$ is an integer. The image is known only from
``projections'' along a few known directions labeled by $j=1, ...,
M$. A direction $j$ is characterized by its unit vector $\bm n^j$,
and its $\pi/2$-rotated vector $\bm t^j$. For each direction $j$,
the projection $\pi^j(y)$, is defined as the line sum of $f$ along
the ``ray'' that projects onto $y$. This ray is the set of points
$\bm x$, such that $\bm x^\top\bm t^j=y$. \revisionter{The
projection is written as
    \be
    \sum_{\bm x^\top\bm t^j=y} f(\bm x)=\pi^j(y)
    \ee
where $|y|\le N/2$.}  The $M$ projections $\pi^j(y)$ are given and
the image $f$ is to be reconstructed.

The total number of pixels is of order $(\pi/4)N^2$. In practice
it is often more convenient to consider all pixels with the
$[-N/2;N/2]^2$ square domain that contain the disk $\cal D$. In
that case, $i$ ranges from 1 to $N^2$, but for those pixels,
$\bm x_i$ which are more distant than $N/2$ from the
origin, $f(\bm x_i)=0$. It is convenient to introduce a vector
representation of the discrete image through the notation
${\mathbf f}=\{f_i\}$ where $f_i=f(\bm x_i)$.  In the sequel, this
vector representation will be used systematically, and denoted (as
well as matrices) by bold characters.

Let us note that in the literature, the sampling of
$\pi^j(y)$ is often assumed to be resolved at the scale of the
separation between the projection of the discrete pixel centers
$\bm x_i$~(see \eg \cite{Batenburg07}). For instance,
when the projection is performed at $\pi/4$ with respect to the
principal axis, pixels are projected onto positions separated by
$1/\sqrt{2}$ and hence the number of projection data is
$\sqrt{2}N$, as compared to $N$ when the projection direction is 0
or $\pi/2$. As the number of projections, $M$, increases, the size
of the discrete vector $\pi^j$ increases, so that the information
content increases quickly.

For that reason, a different discretization choice is
made herein, namely, for any direction, $j$, the $\pi^j(y)$
function is binned over intervals of size 1. Introducing equally
spaced discrete coordinates $y_k$ (with a unit spacing), the
projection $\pi^j(y)$ is transformed into the series of discrete
components
    \be\ba{ll}
    \pi^j_k&\displaystyle=\sum_{|y-y_k|<1/2} \pi^j(y)\\
    &\displaystyle=\sum_i  H(1/2-|\bm x_i^\top\bm t^j-y_k|) f_i
    \ea\ee
where $H$ denotes the Heaviside function, \ie
$H(x)=1$ iff $x\ge0$, and $H(x)=0$ else. Thus the projection data
are collected into $M$ vectors $\bm \pi^j$ for $1\le j\le M$; each
vector $\bm \pi^j=\{\pi^j_k\}$ being of length $N$, $k=1, ..., N$.
In that case, any direction brings approximately the same amount
of information (\ie same number $N$ of equations).

For the sake of simplicity, for any projection
direction, the choice is made to associate each pixel $\bm x_i$
with a unique detector site $y_k$ onto which the projection is
considered (\ie closest integer value as above defined). It is to
be noted that other choices could have been made.  In particular,
a pixel could be partitioned into different rays with weight
corresponding to the area of the square pixel swept by the ray of
unit width.   The first choice is the simplest as the weight
itself is binary.  The second is more realistic when considering
an actual experiment.  In the following only the first choice was made.

In discrete form, these projections are recast in a linear form
    \be
    {\mathbf W}  {\mathbf f}= {\bm \pi}
    \ee
where the projection operator ${\mathbf W}$ is itself a binary
matrix as a result of the above projection discretization. This
linear system is the collection of the projection
equations along different orientations. For a specific direction
$\bm n^j$, this projection operator is denoted by ${\mathbf W}^j$
and second member ${\bm \pi}^j$, so that ${W}^j_{mn}f_n=\pi^j_m$
where $1\le j\le M$, $1\le m\le N$ and $1\le n\le N^2$.

The associated {\em backprojection operator} ${\mathbf B}^j$ is
simply deduced from the transpose of ${\mathbf W}^j$ through a
normalization by the number of pixels $N^j_k$ being projected onto
the same detector site.  The latter is obtained by the projection
${\mathbf W}^j \mathbf 1$ where $1_n=1$ for all $1\le
n\le N^2$, $N^j_k=\sum_n W^j_{kn}$. Let us introduce the diagonal
operator $\mathbf D$ such that
    \be
    D^j_{kl}=\delta_{kl}/N^j_k=\delta_{kl}/\sum_{n=1}^{N^2} W^j_{kn}
    \ee
where $1\le j\le M$, $1\le k\le N$, $1\le l\le N$, and
$\delta_{kl}$ denotes the Kronecker symbol. The backprojection
operator is written as
    \be
    B^j_{nm}= \frac{W^j_{mn}}{\sum_{p=1}^{N^2} W^j_{mp}}
    \ee
for $1\le j\le M$, $1\le m\le N$ and $1\le n\le N^2$, or
equivalently
    \be
    {\mathbf B}^j=(\mathbf W^j)^\top {\mathbf D}^j
    \ee
so that ${\mathbf W}^j {\mathbf B}^j= {\mathbf I}$.

It is to be emphasized that the number of available equations from
the projections is far less than the number of unknowns. Typically
for $N=300$ and $M=4$, the ratio of equations over unknown amounts
to about 1.5~\%, emphasizing the crucial role to be played by the
regularization in the solution.

\section{Algorithm} \label{Sec:Algo}

Most binary reconstruction methods proposed in the
literature, whatever their specific strategy, are iterative.
Hence, the first step of any algorithm is to propose a trial image
for $f$ to initialize the procedure. The obvious route to follow
is to apply the classical fast algorithm known for continuous
reconstruction, and hence, typically, it is proposed to start with
the standard FBP algorithm from the known projections. Although
natural, this procedure is not optimal, and the following
subsection aims at revisiting this first step through a more
genuine estimate of the local probabilities for a site $i$ to be
valued $f_i=0$ or 1.

\subsection{Initialization}\label{ssec:proba}

Let us consider a specific site, $\bm x_i$, and two arbitrary
projection directions, 1 and 2. The corresponding projections of
$\bm x_i$ on the detector line are denoted as $k_1$ and $k_2$
respectively ($W^1_{k_1 i}=1$ and $W^2_{k_2 i}=1$). From the first
direction, and without additional information, the probability
that $f_i=1$ is $p_1$, which is equal to the projection
$n_1=\pi^1_{k_1}$ divided by the total number of pixels
$N_1=\sum_m W^1_{k_1 m}$ along the line. Similarly for the second
direction, the probability is $p_2=n_2/N_2=\pi^2_{k_2}/\sum_m
W^2_{k_2 m}$. The question to be addressed is to provide the
``best'' estimate, $\overline p$, for the probability that
$f_i=1$, knowing $p_1$ and $p_2$, assuming the influence of other
pixels to be negligible. Let us concentrate hereafter
on non trivial cases, where $N_i>1$ and $0<n_i<N_i$ for $i=1$ or
2.

Let us count the number of configurations that are consistent with
$f_i=1$
    \be
    n_+={N_1-1  \choose n_1-1}.{N_2-1  \choose n_2-1}
    \ee
and those corresponding to the alternative choice $f_i=0$
    \be
    n_-={N_1-1  \choose n_1}.{N_2-1  \choose n_2}
    \ee
Hence, using the short-hand notation $q_i=1-p_i$ ($i=1$ or 2), the
ratio of these two quantities is found for all $N_i$ and $n_i$
    \be
    \frac{n_+}{n_-}=\frac{n_1n_2}{(N_1-n_1)(N_2-n_2)}=\frac{p_1p_2}{q_1q_2}
    \ee

\begin{figure}[ht!]
\begin{center}
\includegraphics[width=.9\columnwidth]{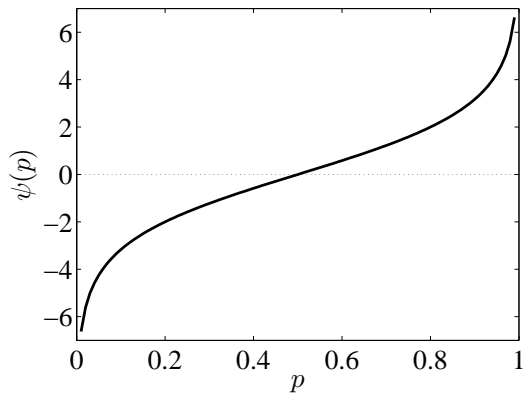}
\end{center}
\caption{ Graph of function $\psi(p)$ when $0 \le p
\le 1$.}\label{fig:psi}
\end{figure}

Without further information, all configurations are considered as
equiprobable and hence, $\overline p$, may be evaluated from the
ratio
    \be
    {\overline p}=\frac{n_+}{n_++n_-}=\frac{p_1p_2}{p_1p_2+q_1q_2}
    \ee
The above combination of the two probabilities to estimate
$\overline p$ appears as quite intricate.  However, if the
following function is introduced
    \be
    \psi(p)=\log(p/(1-p))
    \ee
and the notations $\psi_1=\psi(p_1)$, $\psi_2=\psi(p_2)$ and
$\overline \psi=\psi(\overline p)$, then the above relationship
reads
    \be
    \overline \psi=\log\left(\frac{p_1p_2}{q_1q_2}\right)=\psi_1+\psi_2
    \ee
Figure~\ref{fig:psi} shows function $\psi(p)$.  The divergences at
$p=0$ or 1 are the counterparts of the observation that if a line
is seen as empty or full, then whatever the other projection
values, the pixel (and the entire projection line) is surely
determined. Reverting to probabilities $p$ from $\psi$ is
straightforward
    \be
    p=\frac{e^\psi}{1+e^\psi}
    =\frac{1+\tanh(\psi/2)}{2}
    \ee
In practice for numerical purposes, function $\psi$ is truncated
for arguments close to 0 or 1. A small parameter $\epsilon$ is
introduced and for arguments $x$ less than $\epsilon$ or greater
than $1-\epsilon$, their $\psi$ values are turned into
$\psi(\epsilon)$ or $\psi(1-\epsilon)$ respectively. For all the
examples shown hereafter, a value of $\epsilon=10^{-6}$ was
chosen. It is worth noting that function $\psi$ is the opposite of
derivative of the (Fermi-Dirac) entropy
$S=-p\log(p)-(1-p)\log(1-p)$ with respect to $p$. Exploitation of
this observation in the context of image reconstruction (and more
generally of deconvolution) was proposed by Byrne~\cite{Byrne98}.
In the latter reference, an iterative scheme was proposed to deal
with images obeying constraints such that $0\le f_i\le 1$ for all
$i$ (rather than the binary constraint considered herein), and the
above entropy functional was {\em postulated} as a convenient way
to favor intermediate values within the allowed interval.  This is
to be contrasted with the above derivation based on actual
probabilities.

The additive property of $s=\psi(p)$ is easily generalized to an
arbitrary number of projections.  This observation is the heart of
the proposed algorithm, and in particular the initialization step.
For a given direction, $j$, the projections $\pi^j_k$ are
transformed into probabilities $\pi^j_k/N^j_k$ where $N^j_k=\sum_n
W^j_{kn}$, and further into $s^j_k=\psi(\pi^j_k/N^j_k)$. The
latter is backprojected over the image to be reconstructed. The
summation of these backprojections at each site $i$
    \be\label{eq:aaa}
    \sigma_i=\sum_{j,k} {W}^j_{ki} s^j_k
    \ee
or, $\bm \sigma={\mathbf W}^\top {\mathbf s}$, provides a
resulting ${\bm \sigma}$ vector that is finally transformed back
into probabilities ${\mathbf p}$. This transform reads
    \be\label{eq:initial}
    {\mathbf p}=\psi^{-1}\left({\mathbf W}^\top \psi({\mathbf D}.{\bm \pi}) \right)
    \ee
where the $t$ superscript denotes transposition.  Thus, up to the
$\bm \pi$-$\mathbf s$ substitution, the canonical backprojection
algorithm is to be used.  If $\psi$ is turned into identity, the
above expression reduces to a simple backprojection.

The classical FBP is based on a similar backprojection, but with
the filtered signal $\bm\pi$ (\ie convoluted by the inverse
Fourier transform of the absolute value of the wavevector, $|k|$).
This form of filter results from compensating the spreading of the
back-projected values, prominent for small angles. The algebraic
form of the filter is derived for a continuous distribution of
angles. Let us note that the filtering could also be performed
through a deconvolution on the backprojected unfiltered projection
data.  A similar problem of over-counting is met for a large
number of projections because the same blocks of pixels appear for
neighboring projection directions.  The kernel to be used in the
deconvolution varies with the number of projections, from the
inverse Fourier of $|k|$  for a continuous distribution of angles
to a Dirac distribution for two orthogonal distributions. When the
maximum number of angles does not exceed 10 or 15, the filtering
can simply be ignored.

As a conclusion of this section, two routes could be considered:
\begin{itemize}
\item The first one would consist of i) transforming the
projection with the $\psi$ function, ii) applying a classical FBP
algorithm from this     transformed signal, iii) filtering the
reconstructed image to account for the errors induced by the
incomplete set of projections, iv) finally transform back the
image using the $\psi^{-1}$ function, and threshold the resulting
image to a value of 1/2. It is to be noted that the final step can
be further simplified since thresholding the
$\psi^{-1}$-transformed image at 1/2 is strictly equivalent to
thresholding the untransformed image at a value of $\psi(1/2)=0$.
This approach requires not too few projection angles, but is
simple and fast. It works for instance in the case of a random
distribution of pixels provided the fraction of 1-valued pixels
matches constraints Eq.~\ref{eq:pc_Donoho} as theoretically
obtained by Donoho and Tanner~\cite{Donoho10a,Donoho10b}. Although
suited to ``Type A'', this approach however does not easily allow
for the incorporation of spatial correlations in the reconstructed
image as required for ``Type B''. It will not be further explored
in the present study.

\item The second option is to follow an approach similar to the
iterative correction reconstruction technique, \ie design an
iterative algorithm that progressively corrects the reconstructed
image. This approach is expected to be safer and more precise when
the number of projections is very small, as a one step algorithm
is out-of-reach. It is also {\it a priori} more demanding in terms
of computation time and hence algorithmic efficiency is a critical
issue. Because this approach proceeds by successive corrections,
the first approach (without deconvolution) can be used to provide
an initialization of the reconstructed image.
\end{itemize}

\subsection{Correction}

Before describing the correction step, let us briefly recall a
(block-iterative) ART implementation (SART in the terminology of
Ref.~\cite{Kak01}, \ie simultaneous correction of the image for
each block of projection equations corresponding to a given
direction, $j$, and sequential inspection of the different
directions) to highlight the parallel with the proposed approach.
This algorithm is based on a progressive correction of a current
estimate, ${\mathbf f}^{(n)}$ at step $n$. The projection error
${\bm \pi}^j-{\mathbf W}^j {\mathbf f}^{(n)}$ along direction $j$
is computed and corrected by a uniform translation of $f$ along
each ray (\ie a backprojection) so as to cancel out this error.
All projection directions are successively considered.

One interesting variant is the MART
strategy~\cite{MART72a,MART72b} where M stands for multiplicative.
As for the SART algorithm, a block-iterative formulation, termed
SMART, was introduced~\cite{SMART93,SMART93err,BISMART96}. This
latter strategy consists of {\em scaling} (rather than {\em
translating}) ${\mathbf f}^{(n)}$ by a constant factor along each
ray to meet the projection constraint.  The correction step can
thus be seen as a uniform translation along each ray of
$\log({\mathbf f}^{(n)})$ so that after correction the projection
error exactly cancels out.

Section~\ref{ssec:proba} showed that $\sigma$ is an appropriate
field to apprehend the $f$ image.  In the same spirit as ART and
SMART, the proposed correction step is a uniform increase (or
decrease) along each ray of ${\bm \sigma}^{(n)}$ values so that
the projection constraint along that direction is exactly met. The
latter increment is to be evaluated after a binarization step to
produce an image ${\mathbf f}^{(n+1)}=H({\bm \sigma}^{(n+1)})$.
Hence, the computation of the increment per ray is a little more
demanding than in the two above SART and SMART cases, yet it can
conveniently be achieved via a sorting algorithm when $\mathbf W$
is binary as we now discuss.

Considering a specific projection direction, $j$, and line
reaching the detector at position $k$ in the above procedure, one
may compute the value ${r}_k^j$ that should be subtracted to
$\sigma$ so that the natural binarization of $\bm\sigma-\mathbf
B^j{r}_k^j$ would lead to the known line-sum $\pi_k^j$, or when
considering all lines indexed by $k$ in a vector notation, along
the same direction $j$,
    \be\label{eq:updatesigma}
    {\mathbf W}^j H(\bm \sigma-\mathbf B^j{\mathbf r}^j)={\bm \pi}^j
    \ee
The solution to this non-linear equation is given by selecting the
$\pi_k^j$ largest $\sigma$ values along each line $k$, as one can
consider that the largest values are the ones that are the most
likely to be 1-valued.  Let $\sigma^{*j}_k$ denote the smallest of
them.  Alternatively, one could also consider the ${\cal
N}^j_k-\pi^j_k$ smallest $\sigma$ values, and denote by
$\sigma^{**j}_k$ the largest of them.  Then any value of $r^j_k$
in the interval $\sigma^{**j}_k<r^j_k<\sigma^{*j}_k$ would fulfill
condition~(\ref{eq:updatesigma}).  Thus it is proposed to opt for
the mid-value
    \be
    r^j_k=\frac{\sigma^{*j}_k+\sigma^{**j}_k}{2}
    \ee
For a given projection direction, the correction of $\sigma$ along
each ray is easily performed by a 1D sorting step, or more
precisely by looking for the $\pi$-quantile, for which efficient
algorithms exist~\cite{Quantile}. The resulting corrected image if
binarized would fulfill the appropriate projection for the
considered direction.  However, presumably the projection along
other directions is not correct, and thus the algorithm consists
of successively considering all directions. This step is repeated
twice.

Let us stress that the sorting algorithm to determine the
increment $\mathbf r$ is specific to the case of a binary
projection matrix ${\mathbf W}$ so that rays are decoupled from
each other.  In the case of a more general projection matrix, then
a similar correction step could be performed but the computation
of the increment $\mathbf r$ would be more involved, and hence
less efficient.  This extension has not been explored further.

The proposed algorithm shares the same property as the SMART
algorithm close to 0 where $\psi(x)\sim_{x\to 0} \log(x)$.
However, it also introduces a similar behavior close to $p=1$. The
price to pay is that the correction is a nonlinear problem that
can be solved by a sorting scheme. It is also observed that when
$p$ is close to 0 or 1, a translation of $\psi$ values has a very
small impact on $p$. In contrast for $\psi$ close to 0 (\ie $p$
close to 1/2) the relative influence of a $\psi$-correction is the
largest.

Even though the sought image $\mathbf f$ is binary, it is
important to allow for intermediate values in the course of its
determination. It enables to move continuously from one state to
the other without having to resort to a combinatorial treatment,
or being trapped within a fixed topology. The image $\mathbf f$,
or its $\psi$-transform $\bm \sigma$, are natural quantities to
deal with.

\subsection{Regularization}
\label{ssec:regul2}


Regularization is not introduced here as an extra functional to be
minimized together with the projection constraints.  Rather a
specific iterative procedure is introduced aiming for an image $f$
lying in the constraint set, and stopping once this constraint is
fulfilled.  Because the solution of $\bf W\bf f=\bm \pi$ is
generally non unique in the herein considered cases, the final
image will depend on the specific procedure followed for its
estimation. In some way, such a strategy can be compared to the
spirit to the approach of ``Binary steering''~\cite{Censor}.  It
consists of interleaving a first reconstruction step, and a second
one that tends to favor either 0 or 1 for a real valued candidate
for the binary image. Neither the reconstruction nor the binary
steering is similar to the procedure proposed in this reference,
but the spirit of alternating these two steps can be considered as
qualitatively similar.

Moreover, in the early stage of the algorithm, in addition to
favoring the vicinity of 0 and 1 as values for the image, high
spatial frequencies will be dampened to bias $f$ toward an image
containing few domains and smooth boundaries between them. Along
with iterations, this high-frequency filter will progressively be
tuned down, hence allowing for finer details to be introduced.
However, the convergence of the algorithm relies on the fact that
such a procedure is able to capture the large scale features
quickly, and hence at each iteration only a small correction will
be needed. In the classification of the different types of
sparsity introduced in the introduction, it is observed that the
last stages of the algorithm will definitely rely on the
assumption that boundaries between domains will be sparse (Type
B), while the first stages require that a coarse-grained image has
few pixels of one ``color'' and thus the coarsened image should
rather be of Type A. Because of this observation, we are not able
to derive an operational definition of ``complexity'' or
``simplicity'' suited to the images considered hereafter, and
hence we are not in a position to state precisely the conditions
under which the proposed heuristic procedure converges. However,
some examples shown below illustrate convergence to a satisfactory
solution even if, according to the criterion of Type A complexity,
no solution should have been accessible. This is due to the fact
that spatial correlations reduce complexity. Provided these
correlations are ``gently'' promoted by the procedure, a binary
image may be reconstructed from few projection data. Hence, the
results obtained herein and those of previous comparable
studies~\cite{Batenburg07} suggest that the exact
reconstructability results of Type A could be extended to other
types of signals when space or time correlations are considered.
In this spirit, a measure of complexity derived from that proposed
by Donoho and Tanner~\cite{Donoho10a,Donoho10b} is tested in
Section~\ref{Sec:Complexity}.

The idea of the implemented regularization is to note that the low
frequency part of the image is robust with respect to the missing
information, whereas the high frequency part is more fragile.  In
order to capture long wavelength modes, it is chosen to resort to
a convolution of the current determination of the image,
$f^{(n)}$, with a Gaussian kernel of characteristic size $a$,
$G_a(\bm x)=1/(2\pi a^2)\exp(-\bm x^2/2a^2)$, $g^{(n)}=G_a\star
f^{(n)}$ where $\star$ stands for a convolution. The convoluted
image can be seen as a local average over a centered domain of
area $a^2$. Hence if the convolution image is close to 0 or 1,
then it is likely that the point is inside a subregion of 0's or
1's respectively. In contrast, at the interface between a 0 and a
1 region, the convolution will be of order 1/2. If $g$ is
substituted to $f$ as the current determination of the image, a
correction step is applied that will mostly affect the boundaries
of the 0 and 1 regions. However, as the procedure progresses, the
reconstructed image is expected to come closer and closer to the
true one, and hence, the weight to be given to the
``regularization'', or a priori information, is progressively
reduced. This is easily performed by reducing the extension of the
Gaussian, $a$ to 1.  In this limit, the convolution leaves
$f^{(n)}$ essentially unchanged. Hence, the regularization
initially focusses on the long wavelength components of the
reconstructed image and leaves interfaces to be determined at a
later stage, and progressively, regularization vanishes.  In the
proposed algorithm, a fixed, \ie prescribed pace is chosen, namely
$a$ is uniform and progressively reduces to 1. Thus the
regularization is not defined as belonging to a specific space,
not to the kernel of specific operator, but rather as a
progressively less and less intrusive action on the proposed
solution, first erasing the high frequency components and then
preserving finer and finer scale details.  The problem to be
solved, namely $Wf = \pi$, $f\in\{0,1\}^{N\times N}$, is not
altered, but the regularization introduced here is a way to give a
hierarchy in the determined information on $f$, giving the
precedence to long wavelength over short ones. This will be shown
on a series of numerical (artificial) test cases, to lead
efficiently to a solution. However, no other evidence than these
encouraging numerical examples is offered.

Because the previous correction step involved a sorting procedure
rather than a mere algebraic evolution for each pixel value, it is
difficult to associate the regularization step with the correction
step, and hence these two operations are performed sequentially.
In the following, a matrix notation is chosen to indicate the
convolution by $G_a(\bm x)$, namely, $g=G_a\star f$ is denoted as
${\mathbf g}={\mathbf G}_a {\mathbf f}$ although in practice this
convolution is performed via fast Fourier transforms. The
regularization step consists of a first binarization, followed by
a convolution with a Gaussian kernel, $G_a$
    \be
    {\mathbf g}={\mathbf G}_a H(\bm \sigma)
    \ee
Note that there is no need to transform $\bm\sigma$ into $\mathbf
f$ through $\psi^{-1}$ for the binarization since $H(\bm
\sigma)=H(\mathbf f-0.5)$. The convolution with a Gaussian kernel
is only active in the neighborhood of interface sites between 0
and 1.  The smoothing of the image produced by the Gaussian
convolution kernel is thus a way to ease boundary adjustments
without impeding nucleation of a small cluster nor loosing the
determined information on other pixels. It plays the role of a
surface tension as it tends to smooth out boundaries.

The width of the Gaussian kernel, $a$, is chosen to be larger in
the first steps of the algorithm, and to decrease progressively as
the iteration number $n$ increases.  The following choice was made
in the implementation
    \be\label{eq:Gauss}
    a=1+\alpha^n(a_0-1)
    \ee
with $a_0$ ranging typically from 2 to 4 and $\alpha$ of order
0.7-0.9. Their values are dependent on the image ``complexity''.
They have been chosen to allow for the quickest convergence for a
series of test cases. Note that their precise value affects
predominantly the convergence rate but not the quality of the
result. Let us stress that in all cases the regularization part
involves only very small scale modifications of the image. When
$a$ approaches 1, the effect of convolution essentially vanishes.
The reason for the reduction of the convolution weight is that the
remaining inconsistencies between the known projection vectors
$\bm \pi$ and the projected estimate ${\mathbf W} {\mathbf f}$
become smaller and smaller. The initial value, $a^0$, and its
decrease rate $\alpha$ are the only adjustable parameters of the
proposed procedure.

To summarize, the meta-code of the proposed procedure is given in
Algorithm~\ref{algo:Main}.  The next section illustrates these
different steps on a test case.

\begin{algorithm}[h!]
 \SetAlgoVlined
 \DontPrintSemicolon
 \SetKwInOut{Input}{Input}
 \SetKwInOut{Output}{Output}
 \Input{Projection $\bm\pi$, Initial $a\gets a_0$}
 {\em Initialization}\;
 $\bm \sigma\gets\bf W^\top\psi(\bf D \bm \pi)$
 \tcc*[r]{Eq.~\ref{eq:aaa}}
 \ForEach{direction $j\gets 1$ \KwTo $M$}{
    $\bm\sigma\gets\bm \sigma+\bf B \bf r $                   \tcc*[r]{Eq.~\ref{eq:updatesigma}}
    }
 \KwResult{$\bf f\gets H(\bm\sigma)$}
 \BlankLine
 {\em Iterative correction}\;
 \While{$\Vert \bm r\Vert >\eta$}{
    $a\gets 1+\alpha(a-1)$                              \tcc*[r]{Eq.~\ref{eq:Gauss}}
    $\bm\sigma\gets \psi\left({\bf G}_a \bf f\right)$\;
    \For{$i\gets 1$ \KwTo $2$}{
        \ForEach{direction $j\gets 1$ \KwTo $M$}{
            $\bm\sigma\gets\bm \sigma+\bf B \bf r $           \tcc*[r]{Eq.~\ref{eq:updatesigma}}
            }
    }
 $\bf f\gets H(\bm\sigma)$}
 \Output{Reconstructed image $\bf f$}\;

\caption{Algorithm consisting of two parts: Initialization,  and iterative correction.  $\eta$ is a small parameter  controlling convergence. }\label{algo:Main}
\end{algorithm}

\section{Illustration of the procedure}
\label{sec-illus}

\begin{figure*}[ht!]
\begin{center}
\subfigure[Reference]{\includegraphics[width=.9\columnwidth]{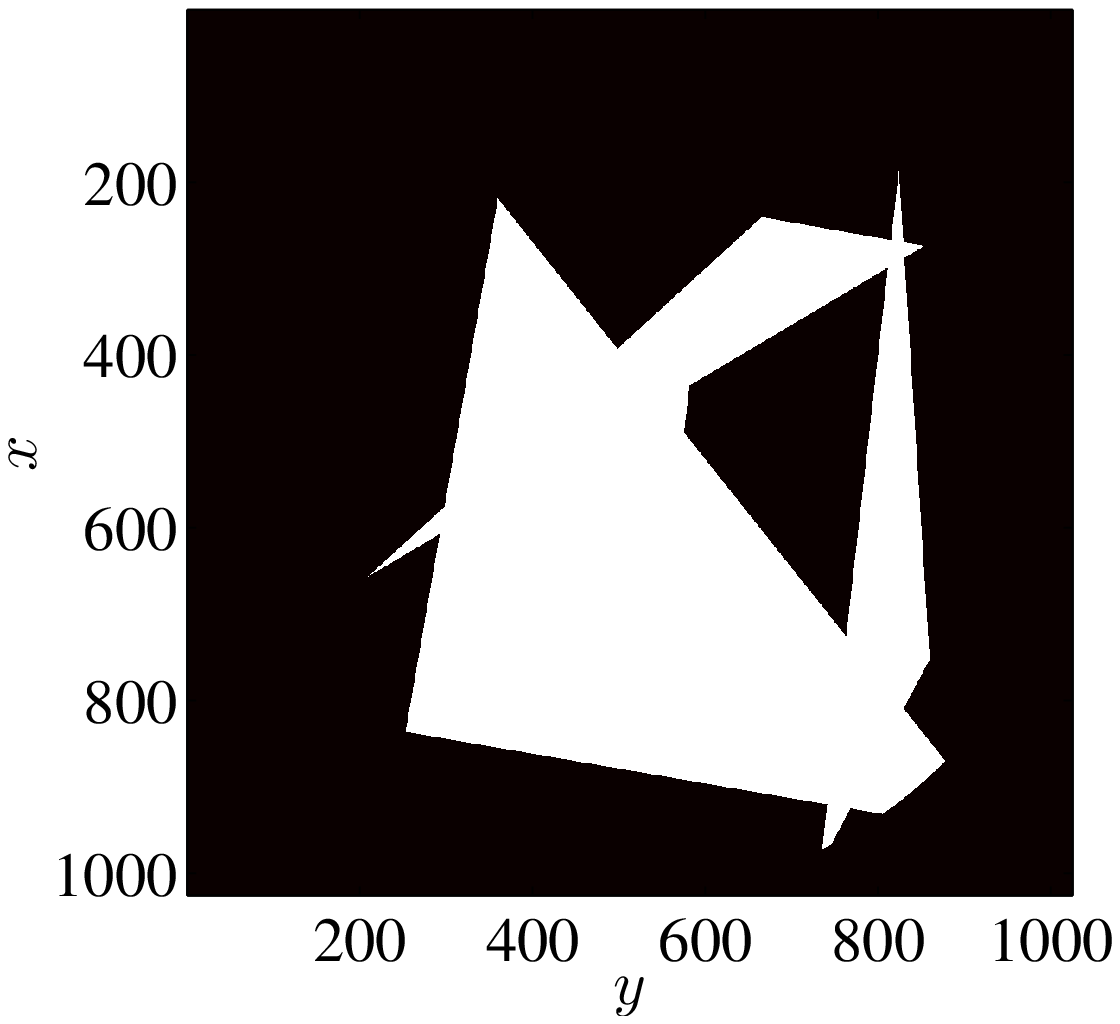}}
\subfigure[$\psi$ backprojection]
{\includegraphics[width=.9\columnwidth]{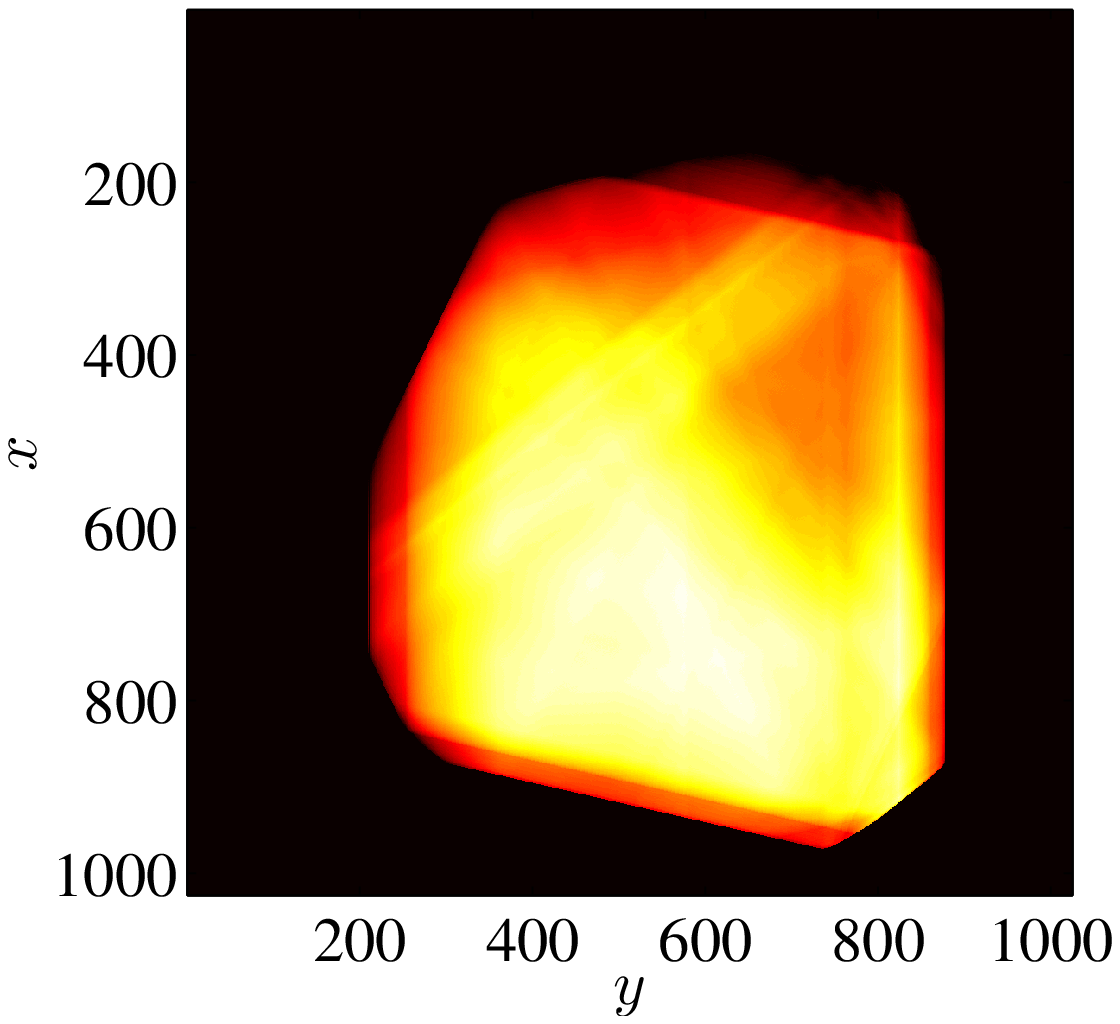}} \subfigure[1st Correction]
{\includegraphics[width=.9\columnwidth]{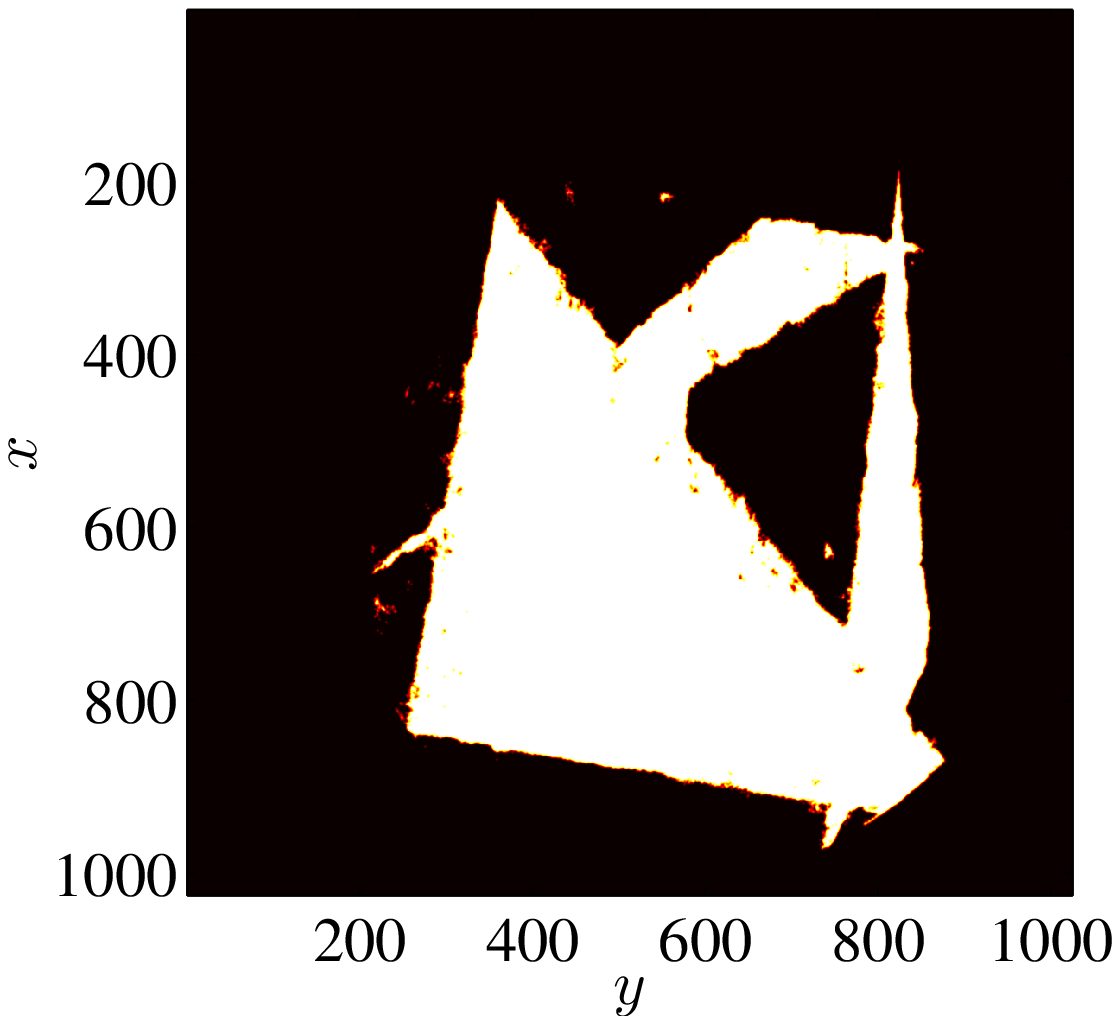}}
\end{center}
\caption{Illustration of the proposed initialization step.  (a) Reference
1-Mpixel image. (b) First raw backprojection, $\psi^{-1}(\sigma)$ based on 7
projections. (c) After a single correction step. The reconstructed image are
shown as a continuous gray (or color) values  from black ($p=0$) to white
($p=1$). Note that the pixel error after binarization of this initialization
step is only 0.82~\%.}\label{fig:PFBP}
\end{figure*}

All the basic ingredients of the algorithm have been defined. The
resulting procedure is now illustrated on a simple example shown
in Figure~\ref{fig:PFBP}a with a rather large definition image (1
Mpixel) as compared to typical definitions from the literature. It
consists of simple shape domains, with however rather sharp
angles, and non convex domains.  In the following the image is
reconstructed from 7 projections (to be compared with the 1600
projections that would be required for a classical
reconstruction).  This small number of projections amounts to
0.45~\% of the classical requirement.

The initialization part consists of: \\
{\sl i)} computing the $\bm \sigma$-image from the backprojection
of $\psi$-transformed probabilities,\\
{\sl ii)} applying a correction step. \\

The resulting images after each of these two steps are shown in
Figures~\ref{fig:PFBP}b and c respectively.  The $\psi$
backprojection produces a $\bm \sigma$ image that captures the
overall shape of the domains in the image, but without access to
small details.  It should be remembered that the $\bm\sigma$ map
is convoluted by a discrete kernel, and no attempt is made here to
perform a deconvolution.  As expected all lines that do not cross
the 1-valued domain have a very small $\sigma$-value.  After a
single correction step, the $\sigma$ values have been adjusted so
that the natural binarization of the corrected image matches
exactly the known projection along the last visited direction. The
resulting image is shown in Figure~\ref{fig:PFBP}c. The latter is
already quite close to the original reference image.  The
difference between the two (fraction of pixels having the wrong
$f$-value) amounts to 0.82~\%.

The remainder of the procedure consists of repeating the
correction step.  The only difference between the successive steps
is the fact that the width of the Gaussian filter progressively
decreases to 1.  In the present example, $a_0=3$, and
$\alpha=0.75$.

Evaluation of the projection error $\Vert{\mathbf W}{\mathbf
f}-\bm \pi\Vert$ is performed, and for the artificial cases
considered in this study, a pixel error is also computed from the
difference $\Vert{\mathbf f}_{est} -{\mathbf f}_{orig}\Vert$. If
the error is less than a threshold value (or a maximum number of
iterations is reached) the code terminates, otherwise a new
correction step is performed starting from the obtained $\mathbf
f$.

In the present case, only four correction steps are needed to
reach an error free solution.  Table~\ref{Tab:Convergence} reports
the relative pixel and projection errors as a function of
iteration number.

\begin{table}
\caption{Convergence rate of the procedure for the example of
Figure~\ref{fig:PFBP}.} \label{Tab:Convergence}
\begin{center}
\begin{tabular} {ccc}
\hline
Iteration &
Relative Pix. Error & Relative Proj. Error\\
& (\%) & (\%)\\
\hline
(init.) & 0.81 & 1.09\\   
1  & 0.23 & 0.42\\        
2  & 0.04 & 0.11\\        
3  & 0.001 & 0.004\\      
4  & 0 & 0\\              
\hline
\end{tabular}
\end{center}
\end{table}

\begin{figure*}[ht!]
\begin{center}
\subfigure[Iteration \# 1] {\includegraphics[width=.9\columnwidth]{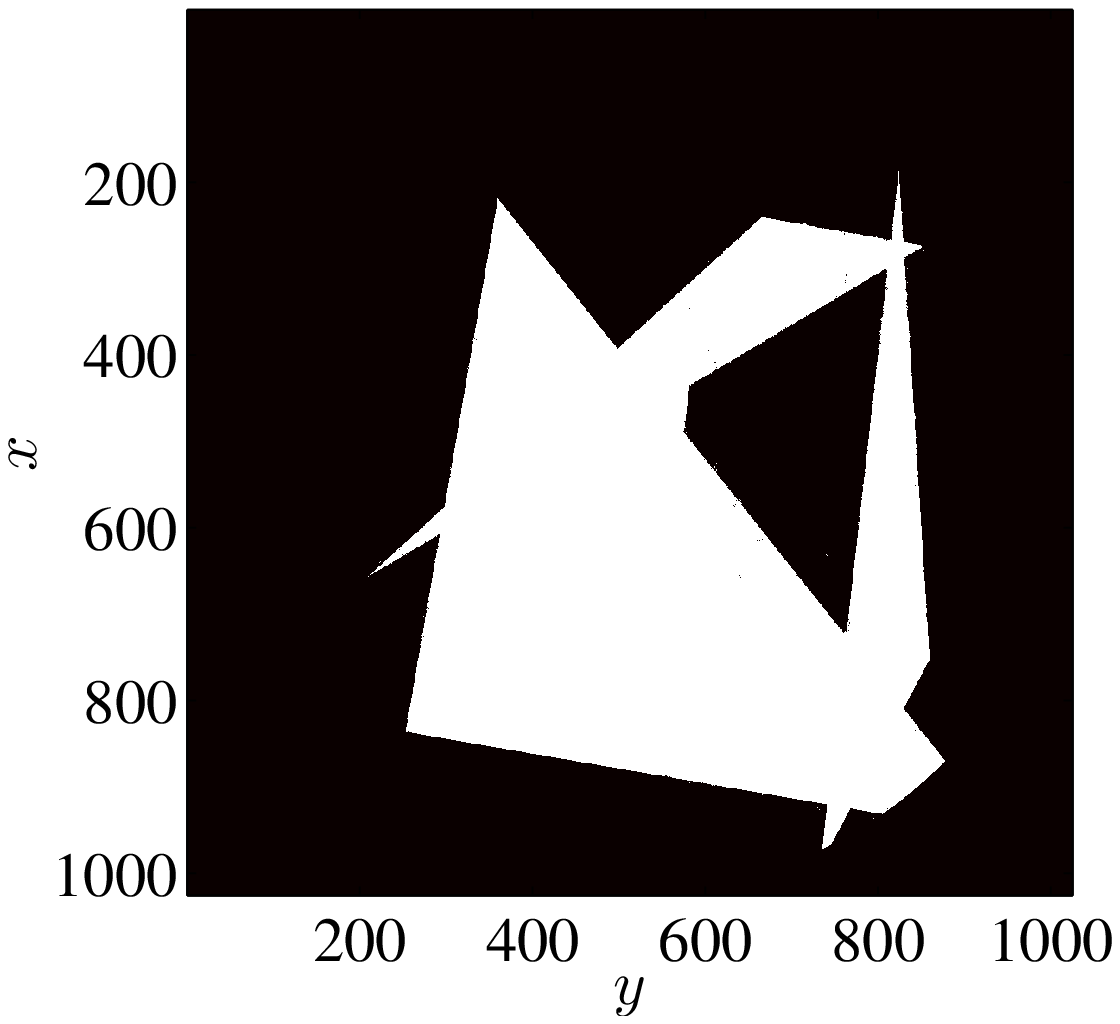}}
\subfigure[Iteration \# 2] {\includegraphics[width=.9\columnwidth]{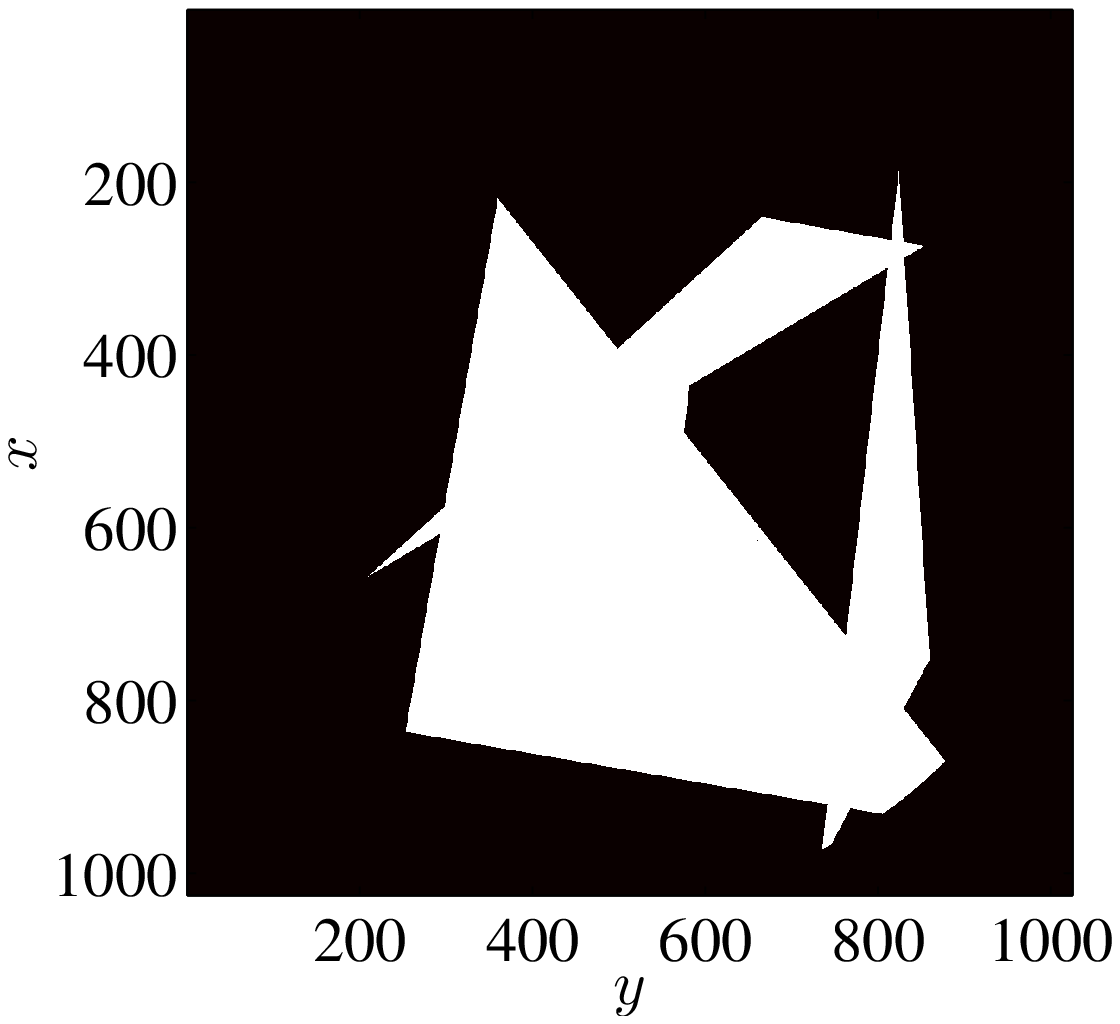}}
\subfigure[Iteration \# 3] {\includegraphics[width=.9\columnwidth]{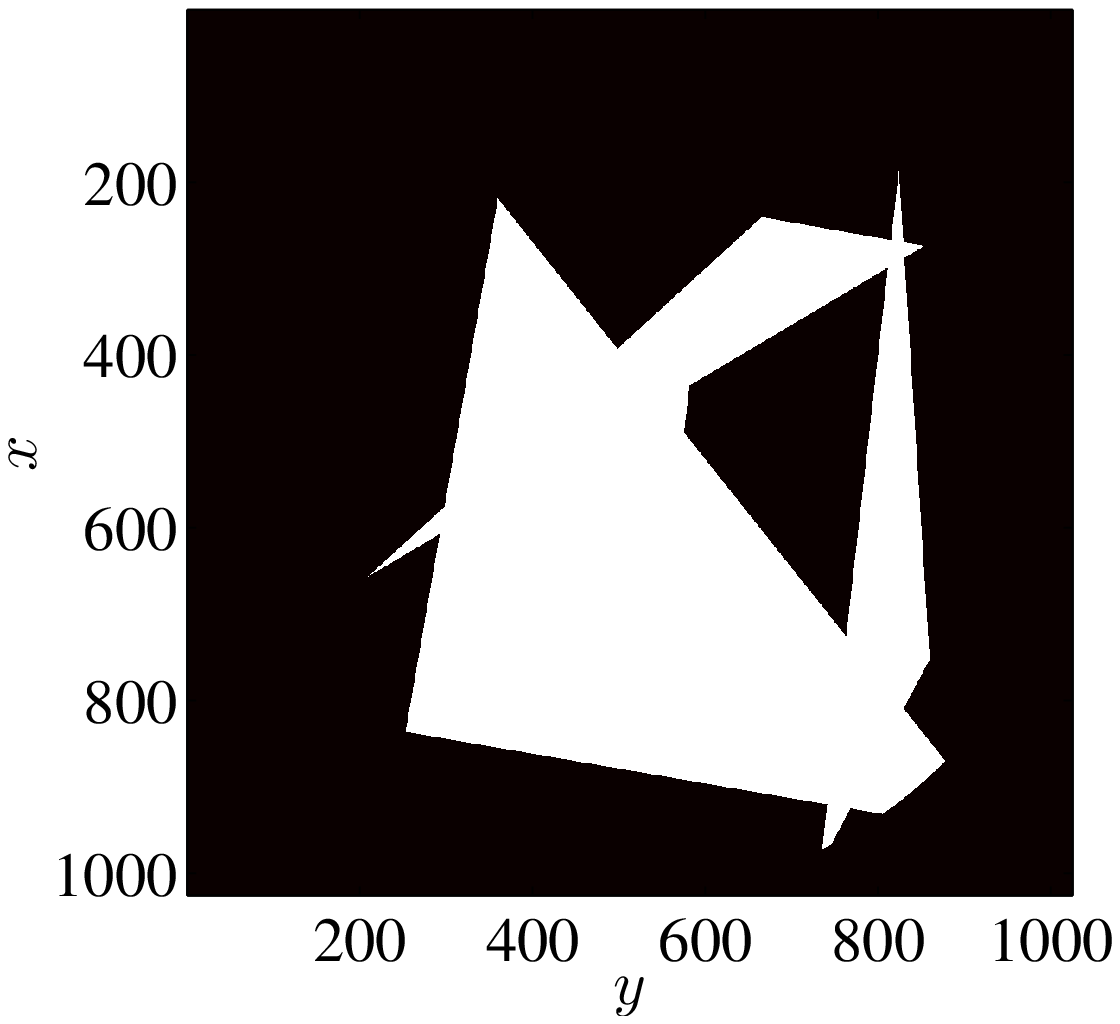}}
\end{center}
\caption{ Different stages of the reconstruction after (a) 1, (b) 2 and (c) 3
iterations. The last image differs from the original model by only 10 pixels
(Figure~\ref{fig:PFBP}(a)).}\label{fig:Reconst}
\end{figure*}

The program was written in Matlab$^{\circledR}$, and run on a
single processor PC (2.3-GHz Core i5 processor) without fancy
optimization. Computation time for this 1-Mpixel image is 8.4 s.

\subsection{Relation with Batenburg's algorithm}

As earlier mentioned, Batenburg proposed a rather efficient
procedure for binary
reconstruction~\cite{Batenburg07,Batenburg08}.  Actually the
spirit of the present approach shares some similarities with that
work. After initialization (performed using a standard FBP
algorithm), the proposed approach by Batenburg is to find
successive binary images that fulfill exactly the projection
constraint along two directions, and to visit successively
different pairs of projection direction.  In addition to
fulfilling two projection constraints, the estimate ${\mathbf
f}^{(n)}$ is chosen to maximize its projection along a predefined
direction ${\mathbf g}$, through the scalar product ${\mathbf
f}^{(n)\top}{\mathbf g}$.  Batenburg's clever observation was to
note that the satisfaction of the two projections could be
rewritten as a max flow/min cut graph optimization problem.  The
maximization of the projection along $\mathbf g$ was then simply
obtained as a simplex-type problem. The resulting procedure was
shown to be more efficient than all available alternative
algorithms.  In fact the present scheme can be seen in a similar
spirit with however some significant differences. First, the
proposed initialization leads to a predetermination of the image
much closer to the sought solution, a very helpful property for
convergence. Second, the proposed correction step considers a {\em
single} direction. The problem is still of simplex type, but so
simple that the solution is nothing but the sorting scheme that
was proposed. This results in a much simpler code.  Note that
Batenburg and Sijbers~\cite{Batenburg09} also considered a single
projection constraint at a time, later on, but this variant was
found to be much less efficient than the original 2-projection
constraint.

\section{Multiscale acceleration}\label{Sec:Multiscale}

As presented so far, the algorithm has been tested in a number of
different cases, always successfully, \ie down to a zero remaining
pixel error (provided enough projections are considered). However,
it is possible to speed up the convergence process using a
multiscale variant. The image $\mathbf f$ to be reconstructed can
be coarse-grained by gathering 2$\times$2 pixels into super-pixels
whose value is determined by a majority rule (and a random choice
if the sum of pixels is equal to 2).  The resulting problem is
simpler in the sense that the projection information $\bm \pi$ is
reduced by a factor of 2, but the number of pixels to be
determined is reduced by a factor of 4.  Hence the ratio of
projection information and the unknowns is doubled. The solution
to this reduced-scale problem can be re-expanded giving to each
constituent pixel the same value as that of the mother
super-pixel.  This image can be used as the initialization of the
relaxation step.  Such a coarsening operation defines one level of
a pyramidal construction that can be repeated up to a chosen
maximum level $n_{level}$ (\eg $n_{level}=5$ reduces the number of
unknown sites by a factor of 1024). Such a multiscale strategy
reduces drastically the required number of relaxation steps and
thus cuts down the computation time by a significant amount
(typical gains of a factor 2 or more are obtained).

\begin{figure}[ht!]
\begin{center}
\includegraphics[width=.9\columnwidth]{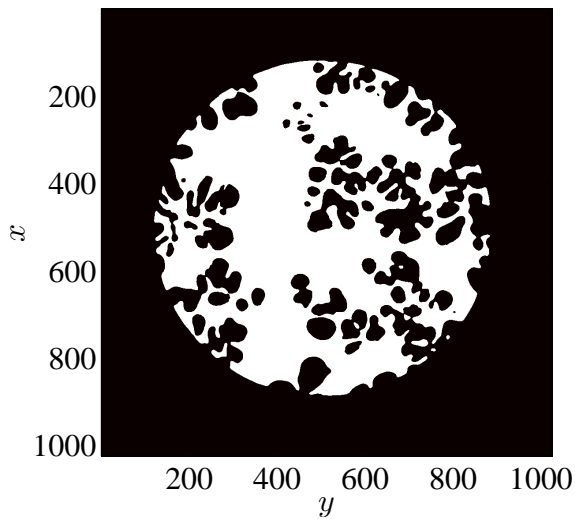}
\end{center}
\caption{ Binarized pattern reproduced from an
actual tomography. No less than 15 projections are needed to
complete an error-free reconstruction using a single scale procedure.
Parameter values are $a_0=4$, $\alpha=0.87$.
(Original tomographic data
were kindly provided by E. Gouillart and C.~Zang~\cite{Zang}.)}\label{fig:AlCu}
\end{figure}

\begin{figure}[ht!]
\begin{center}
\subfigure[Monoscale]
{\includegraphics[width=.9\columnwidth]{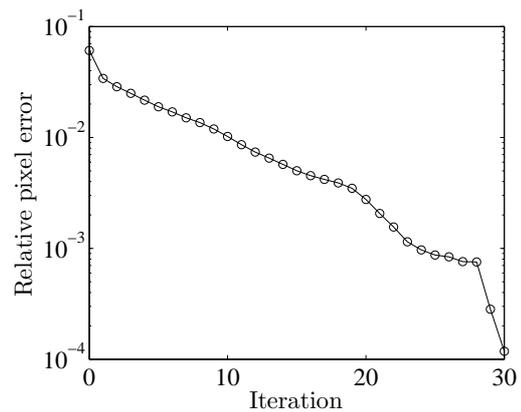}}
\subfigure[Multiscale]
{\includegraphics[width=.9\columnwidth]{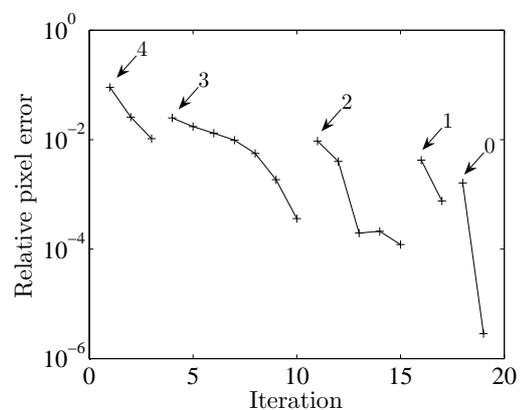}}
\end{center}
\caption{Semi-log plot of the relative pixel
error versus iteration number using the monoscale approach (a)
reaching an error free solution at the
31-st iteration. The 
multiscale case is shown in (b).
At the completion of a level, the
error naturally increases when the coarse image is used as an
initialization for the finer one.  Arrows indicate the start of a
given level.}\label{fig:Error}
\end{figure}

The example shown in Figure~\ref{fig:PFBP} requires too few
iterations for convergence to highlight the benefit of the
multiscale approach.  A more complicated texture binarized from an
actual tomographic reconstructed image, of size
$1025\times1025$~pixels is chosen as a test case (it will be
analyzed below, Figure~\ref{fig:AlCu}). Because of the complex
microstructure, a minimum number of 15 projections is needed to
allow for an exact reconstruction. This should be compared with
the 1500 projections needed for the original reconstruction.
Without multiscale acceleration, 31 iterations are needed,
resulting in a computation cost of 189 s. With 5 levels of the
multiscale procedure, the same exact reconstruction is reached in
28 s \ie about 7 times faster.  The number of iterations from
level 4 to 0 are successively 3, 8, 6, 3 and 3. However due to the
rapid decrease of computation time with the level number, most of
the computation time (more than half of it) is spent at level 0.
The residual relative pixel error as a function of iteration
number is shown in Figure~\ref{fig:Error}. It may also be noted
that using the multiscale procedure, a perfect reconstruction
could be obtained for a smaller number of projections (\ie 13).

\section{Test Cases}\label{Sec:TestCases}

Up to now, only two examples have been shown for illustration
purposes.  A variety of other test cases have been tried with
smooth or angular shapes, equal sized domains or having a wide
distribution of characteristic scales.  In all the cases where the
number of projections is sufficient, a successful (\ie error-free)
reconstruction was achieved, tuning parameters such as $a^0$ or
$\alpha$ in a small range if necessary. However, depending on the
complexity of the patterns, the minimum number of projections
differs.  When the number of projections is too small, the error
saturates at a plateau value.  The number of projections may be
roughly related to the number of interfaces between 0 and 1
pixels, even though high curvatures also play a significant role.
It has been shown mathematically that convex objects only require
at most seven projections to be reconstructed exactly, and that
this number can be reduced to four for specific
directions~\cite{Convex}. Conversely, a complex pattern taken from
a real tomography as shown in Figure~\ref{fig:AlCu} required no
less than 15 (for monoscale) or 13 (for multiscale)  projections
for a perfect reconstruction (for a 1-Mpixel image). An {\it a
priori} evaluation of the ``complexity'' of an image, and the
minimum number of projections needed to reconstruct it, is an
interesting and important question that deserves further studies.

In order to illustrate the performances of the proposed algorithm,
we reproduced a series of tests initially proposed by
Batenburg~\cite{Batenburg07}.  In the latter reference, a detailed
comparison of the algorithm proposed by the author with two of the
most powerful approaches proposed so far.  More precisely, the
first one is the extension proposed by Weber {\it et
al.}~\cite{Weber03}, of the linear programming algorithm of
Fishburn {\it et al.}~\cite{Fishburn97}, and the second one is a
greedy algorithm without any smoothness prior~\cite{Gritzmann00}.

All those benchmark tests are carried out on
$257~\times~257$-``pixel'' lattices, and for each type of pattern,
the numbers indicated in the tables correspond to an average over
200 random samples. Although we tried to remain as close as
possible from the initially proposed references, some adjustments
were necessary. All the objects present in the binary image are to
fit in a circle inscribed in the square.  In doing so, the number
of objects is preserved rather than their density (so that if the
number of 0-1 interfaces is a crucial parameter, complexity is not
altered). For a dilute collection of objects such interface sites
remain the same. A second difference worth mentioning is the fact
that the projections used in the present analysis have a fixed
size equal to the width of a pixel in the binary image. In
contrast, in Ref.~\cite{Batenburg07},  all distinct point-wise
projections were distinguished, so that the actual amount of
information (\ie number of line-sums) may be significantly larger
than the one used in the present case (when the angle tangent does
not coincide with a fraction of low integers).

\subsection{Polygon test cases}

\begin{figure*}[ht!]
\begin{center}
\subfigure[$n=1$, $p=25$]
{\includegraphics[width=.9\columnwidth]{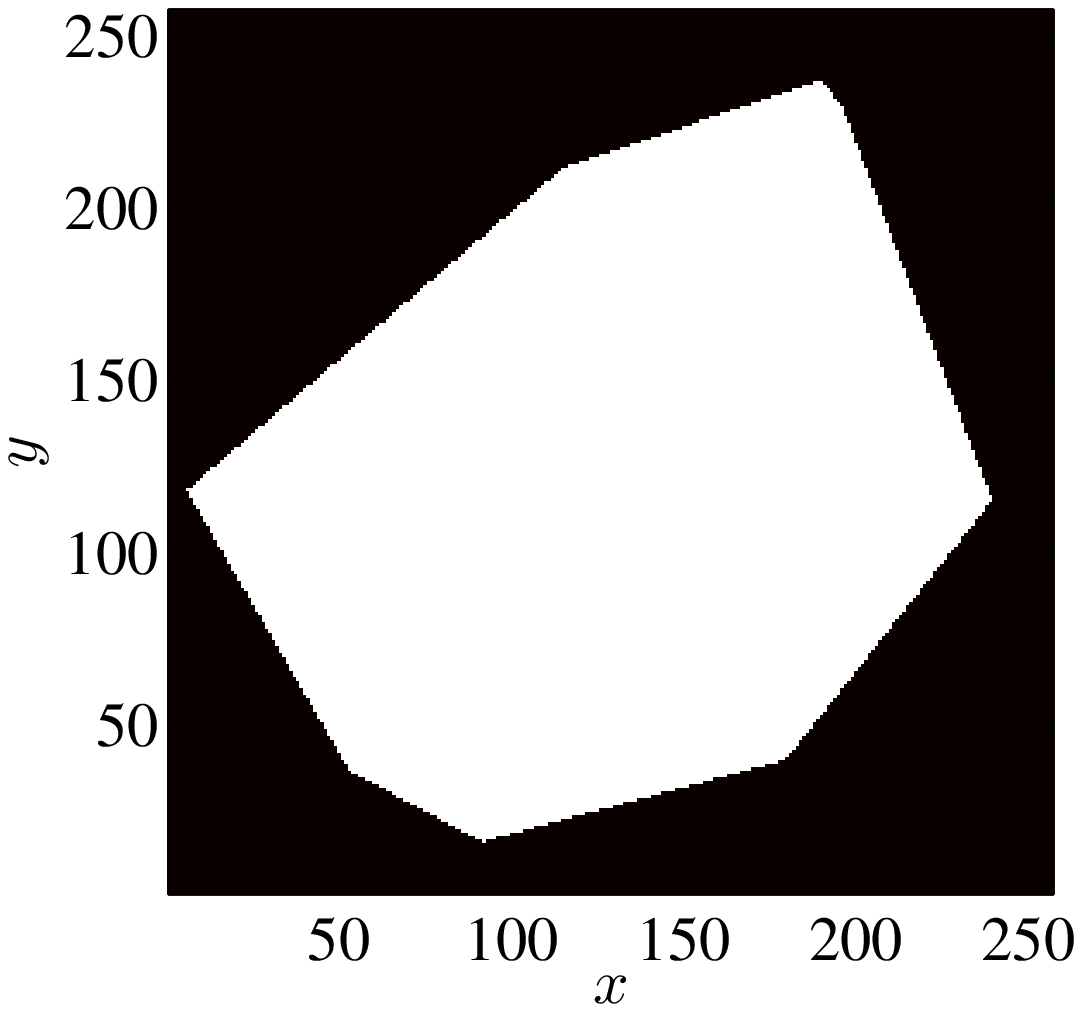}}
\subfigure[$n=5$, $p=8$]
{\includegraphics[width=.9\columnwidth]{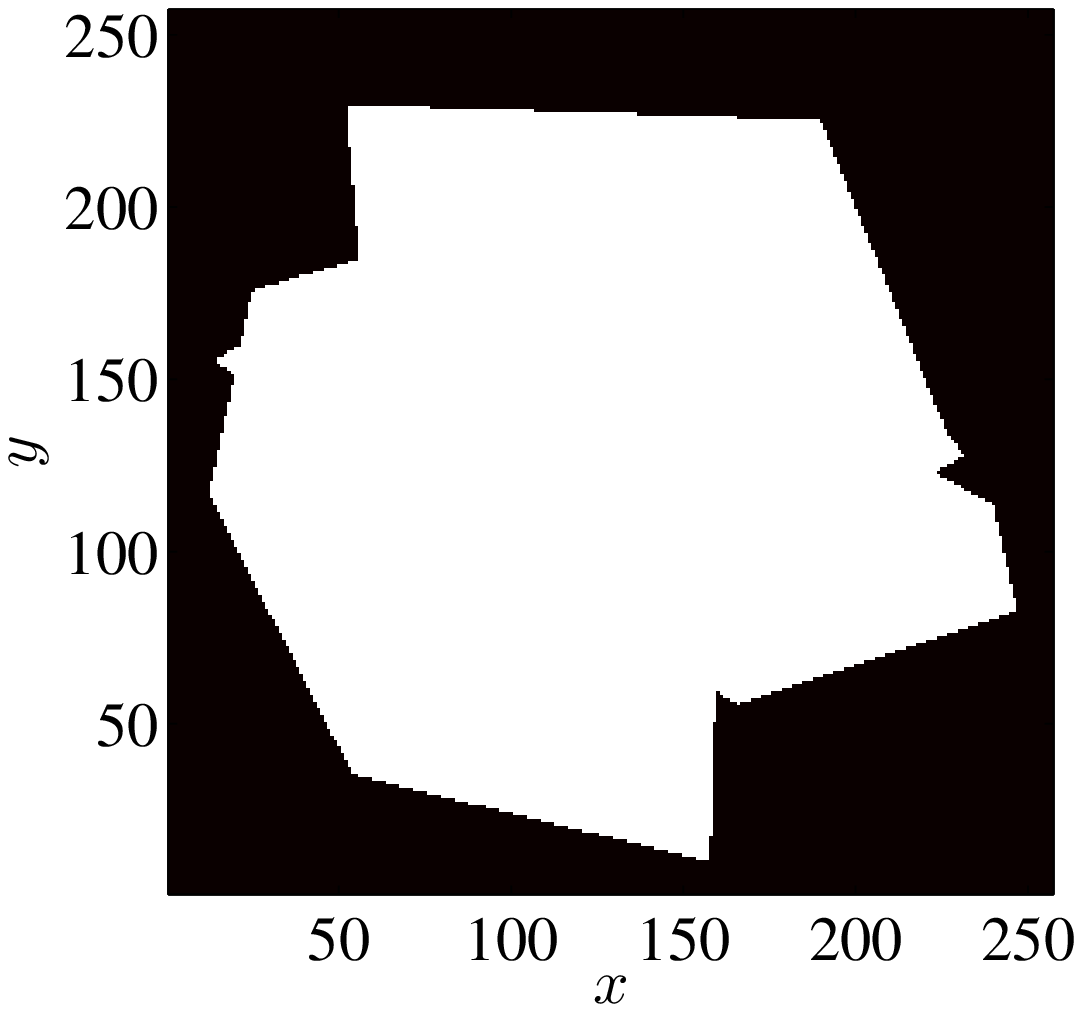}}
\\
\subfigure[$n=12$, $p=4$]
{\includegraphics[width=.9\columnwidth]{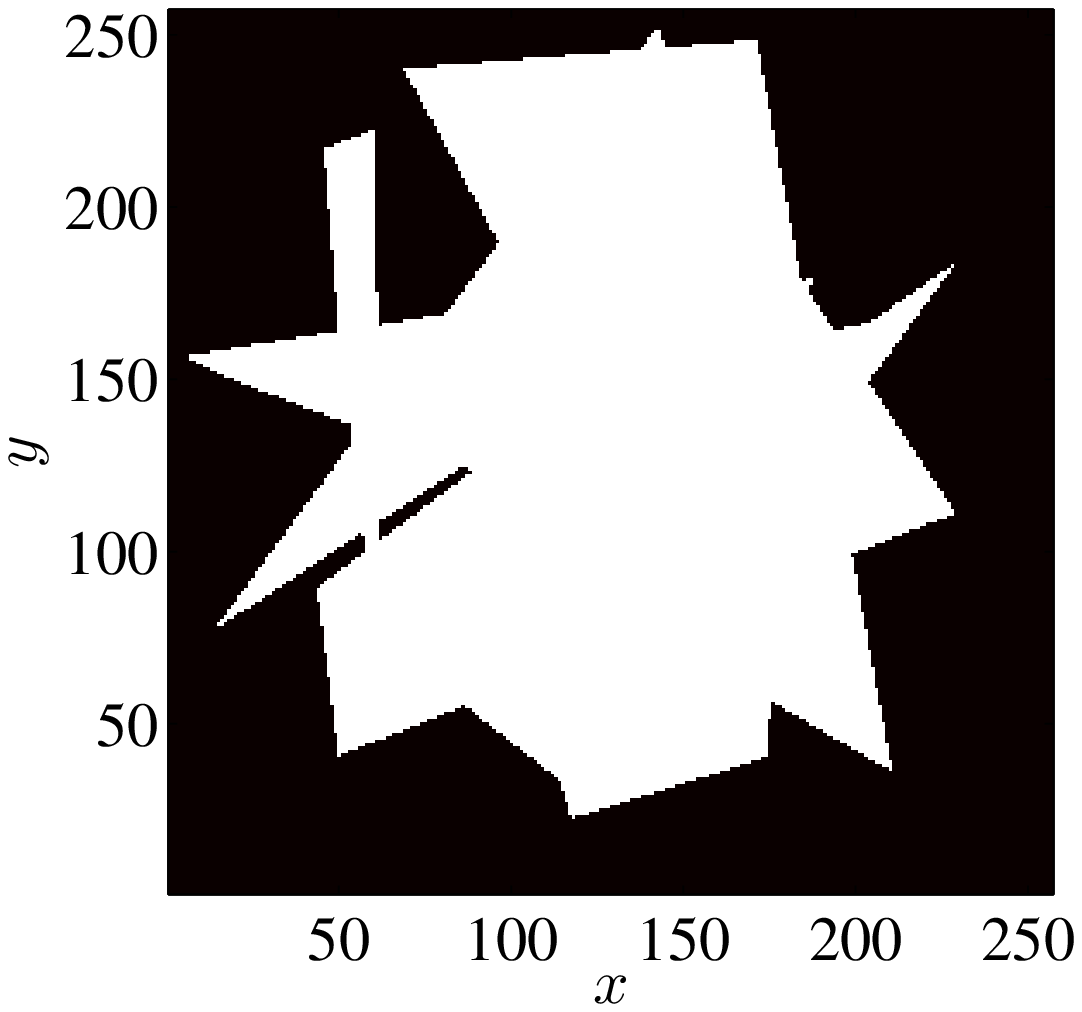}}
\end{center}
\caption{Polygon test cases for the three sets of parameters
$(n,p)$ used. }\label{fig:TestPoly}
\end{figure*}

The first test case was constructed from the union of $n$ convex
polygons.  Each polygon is constructed from a random set of $p$
points within the largest inscribed circle, from which a convex
hull is considered.  The two parameters $(n,p)$ thus characterize
the image.  Three such cases where considered, ranging from a
single $n=1$ convex polygon (with a large number of points $p=25$
although fewer actually are vertices of the convex envelope), to a
large number $n=25$ of such polygons constructed from few points
($p=4$).  The intermediate case consists of $n=5$ polygons built
from $p=8$ points.  The first case is an elementary problem, the
two subsequent examples exhibit details and non convexity that are
much more difficult to capture.  Figure~\ref{fig:TestPoly} shows
an example of the three choices of $(n,p)$ values.

Those examples were then analyzed with a variable number of
projections $M$. For each of them, the proportion of perfect
reconstructions (out of the 200 trials) was recorded together with
the projection error (\ie the sum of absolute value of differences
between original data and reconstructed projections, averaged over
all 200 samples), and the pixel error (\ie the number of pixel
differing in the binary image and its reconstruction, averaged
over all 200 samples). Finally the computation time is recorded.
Note that the same set of parameters was chosen to complete the
analysis, namely, three coarse-graining steps were considered, and
the total number of iterations per scale was limited to a maximum
value of 20 (or less if perfect reconstruction is reached
earlier). Data extracted from Ref.~\cite{Batenburg07} are reported
in Table~\ref{Tab:Polygon_Bat}, whereas the corresponding result
for the present algorithm is given in
Table~\ref{Tab:Polygon_Present}.

\begin{table*}
\begin{center}
\caption{ Polygon test cases reproduced from
Ref.~\cite{Batenburg07}.}\label{Tab:Polygon_Bat}
\begin{tabular}{cccr@{.}lr@{.}lr@{.}lrr}
\hline
$n$ &$p$ &$M$  &\multicolumn{2}{c}{\# perfect (\%)} &
\multicolumn{2}{c}{Proj. error} &\multicolumn{2}{c}{Pix. error}  &Time (s)\\
\hline
1  &25 &3 &93&5        &0&5     &24&0       &13~~~~  \\
&      &4 &~~~~100&0   &0&0     &0&0        &11~~~~  \\
\noalign{\smallskip\smallskip}
5  &8  &3 &29&5        &~~~~32&0  &~~~~538&0  &14~~~~ \\
&      &4 &100&0       &0&0     &0&0        &11~~~~ \\
&      &5 &100&0       &0&0     &0&0        &10~~~~ \\
\noalign{\smallskip\smallskip}
12 &4  &4 &95&0        &12&0    &62&0       &18~~~~ \\
&      &5 &100&0       &0&0     &0&0        &18~~~~ \\
&      &6 &100&0       &0&0     &0&0        &14~~~~ \\
\hline
\end{tabular}
\end{center}
\end{table*}

\begin{table*}
\begin{center}
\caption{ Polygon test cases with the present method (number of levels $=3$,
$a_0=4$, $\alpha=0.87$).} \label{Tab:Polygon_Present}
\begin{tabular}{cclr@{.}lr@{.}lr@{.}lr}
\hline
$n$ &$p$ &$M$  &\multicolumn{2}{c}{\# perfect (\%)} &
\multicolumn{2}{c}{Proj. error} &\multicolumn{2}{c}{Pix. error}  &Time (s)\\
\hline
1   &25    &3    &92&5             &1&0       &3&0     &0.36~~~~ \\   %
&          &4    &~~~~~99&0        &0&0       &0&6     &0.45~~~~ \\   %
\noalign{\smallskip\smallskip}
5    &8    &3    &63&5             &1&0       &1&7     &0.50~~~~ \\   %
&          &4    &99&0             &1&0       &5&7     &0.43~~~~ \\   %
&          &5    &100&0            &0&0       &0&0     &0.43~~~~ \\   %
\noalign{\smallskip\smallskip}
12   &4    &4    &90&0         &~~~~2&0  &~~~~21&0     &1.92~~~~ \\   %
&          &5    &97&5            &1&0        &1&3     &1.31~~~~ \\   %
&          &6    &100&0            &0&0       &0&0     &1.04~~~~ \\   %
\hline
\end{tabular}
\end{center}
\end{table*}

Although the geometry of this test case is simple, the very small
number of projection directions makes the problem difficult.  In
particular, even without projection error, the image may not be
reconstructed perfectly, (see \eg $n=1$, $p=25$, $M=4$ in
Table~\ref{Tab:Polygon_Present}).  The proportion of perfect
reconstructions is comparable in Ref.~\cite{Batenburg07} and with
the proposed algorithm, apart for the $n=5$, $p=8$ and $M=3$ case,
where the present code has about half unperfect reconstructions.
In most other cases the differences are not significant, as the
errors are not weighted in this criterion. When considering the
mean projection error, it is observed that it never exceeds 2 with
the present approach. The mean pixel error is 21 wrong pixels in
the worst case.  It is to be stressed that this represents only an
extremely small fraction, $3 \times 10^{-4}$, of pixels.  Thus in
terms of errors, the proposed approach performs generally better
than Ref.~\cite{Batenburg07}. Finally, in terms of computation
time, we observe a reduction in the computation time by a factor
of order 10 to 20.  Let us stress that this latter comparison is
fragile, as it amounts to compare codes written in different
languages, and run on different computers.  In the present case
however, the code was not optimized, and was run on a standard
laptop computer.

\subsection{Ellipse test case}

\begin{figure*}[ht!]
\begin{center}
\subfigure[$n=15$, $r_{min}=20$, $r_{max}=40$]
{\includegraphics[width=.9\columnwidth]{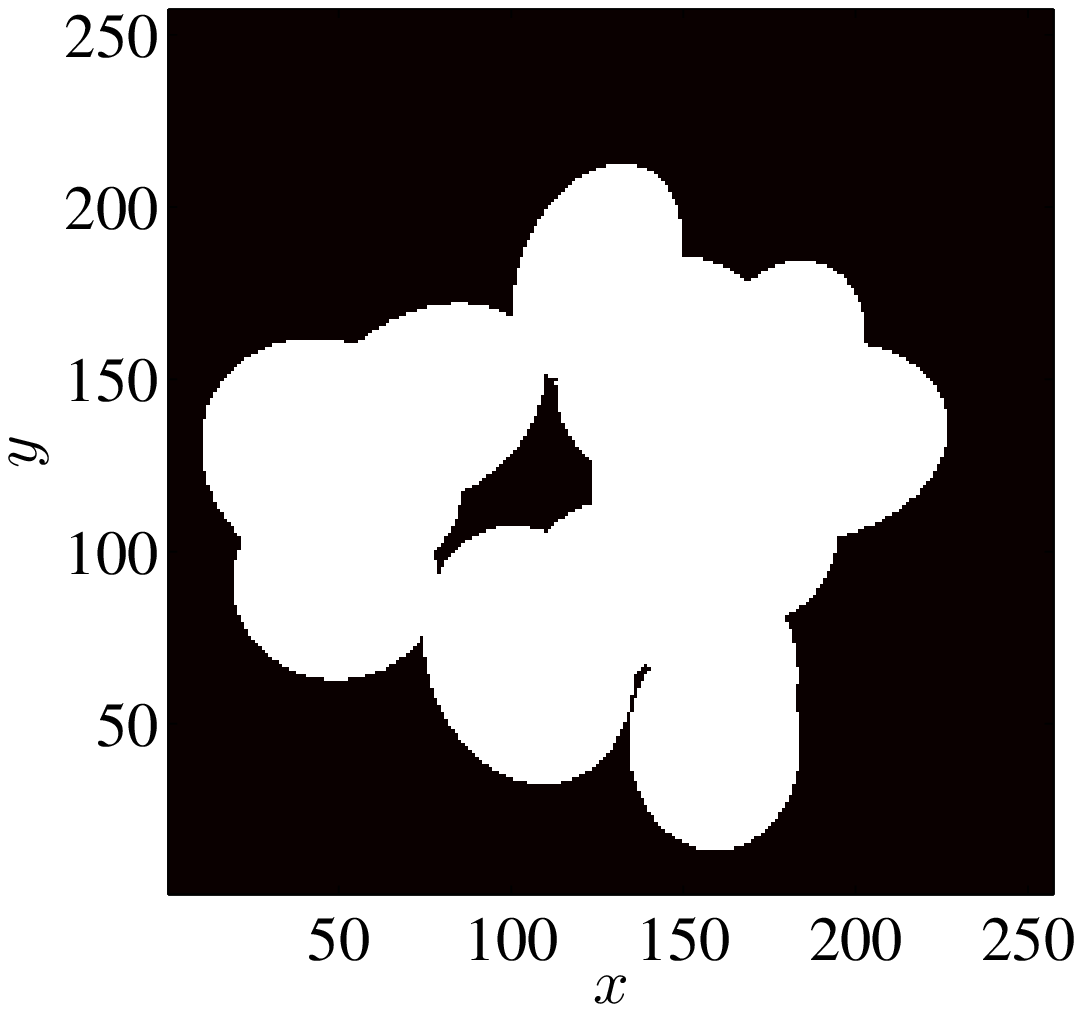}}
\subfigure[$n=50$, $r_{min}=5$, $r_{max}=35$]
{\includegraphics[width=.9\columnwidth]{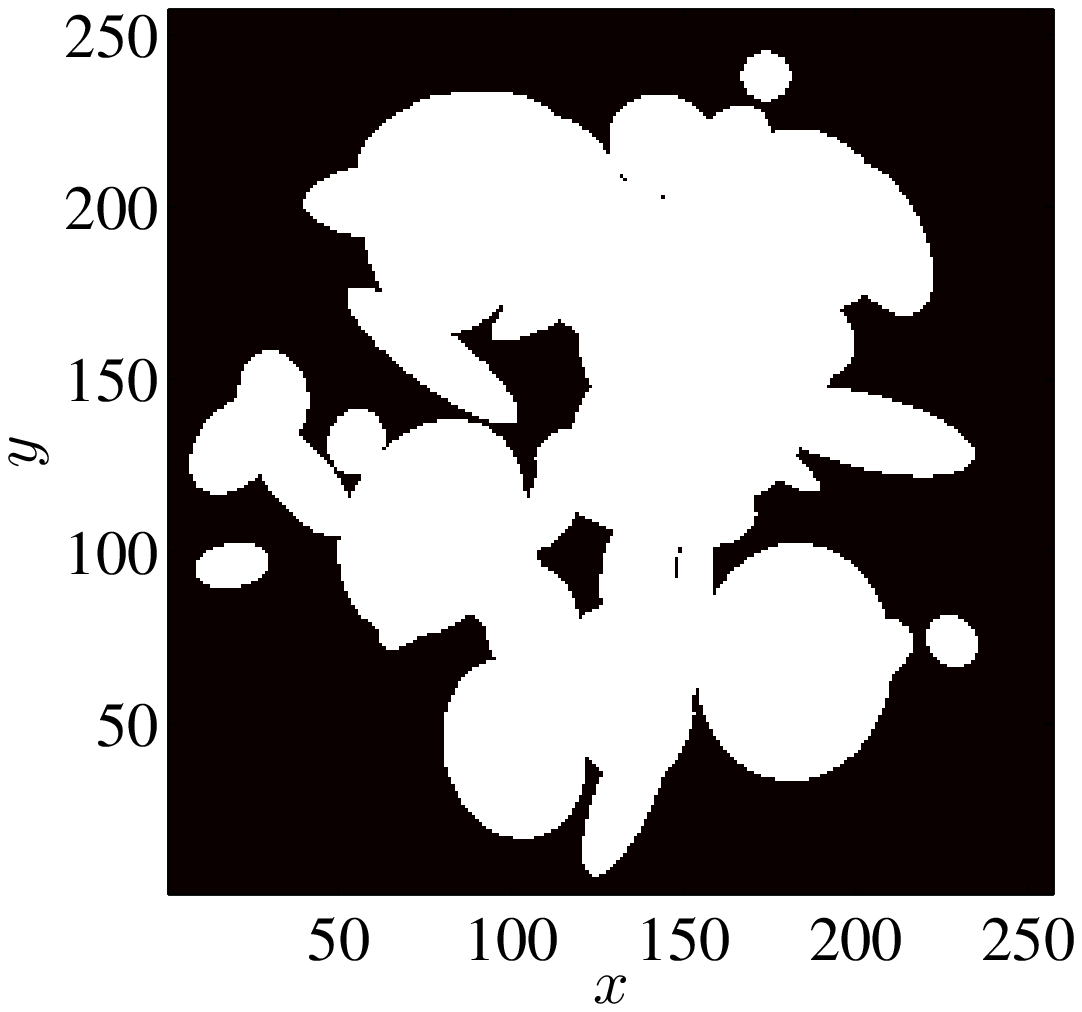}}
\\
\subfigure[$n=50$, $r_{min}=5$, $r_{max}=25$]
{\includegraphics[width=.9\columnwidth]{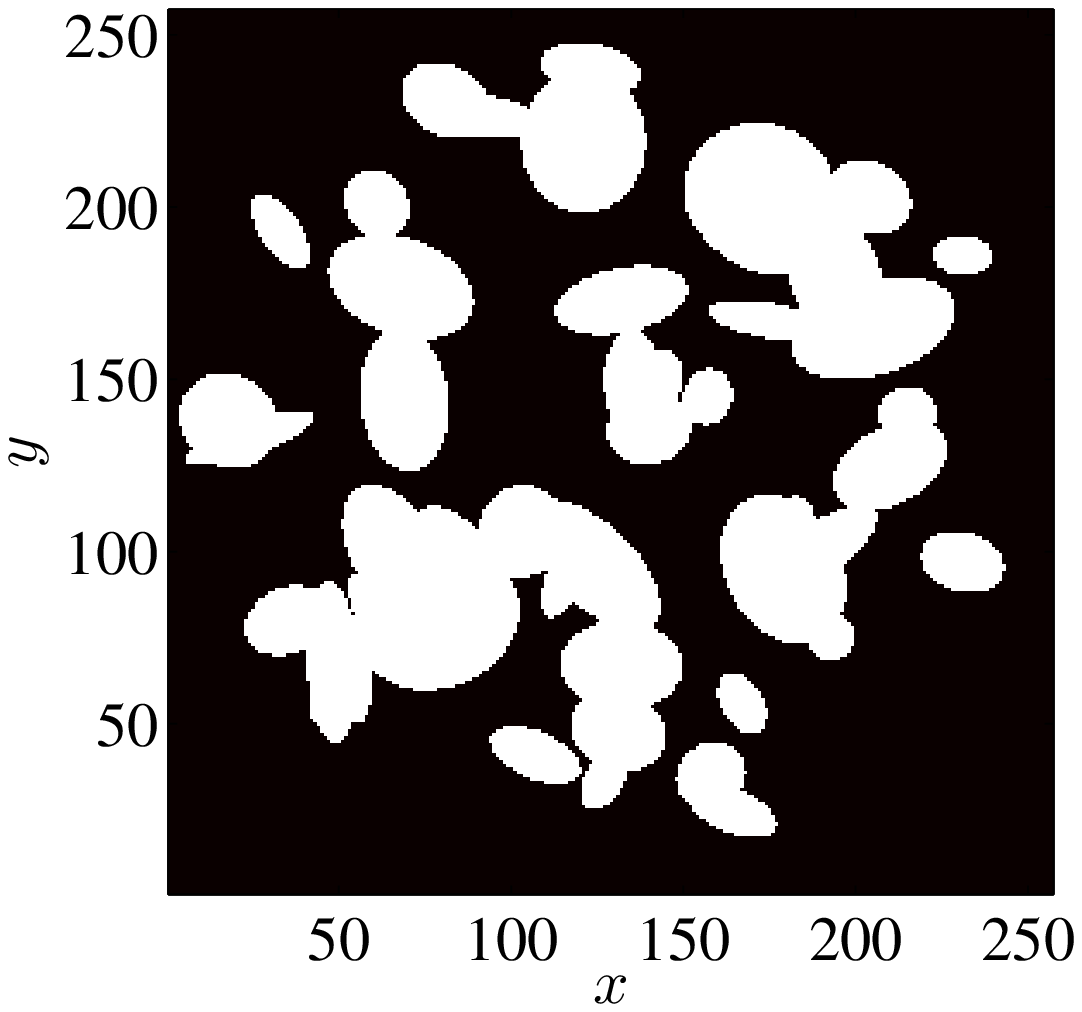}}
\subfigure[$n=100$, $r_{min}=5$, $r_{max}=25$]
{\includegraphics[width=.9\columnwidth]{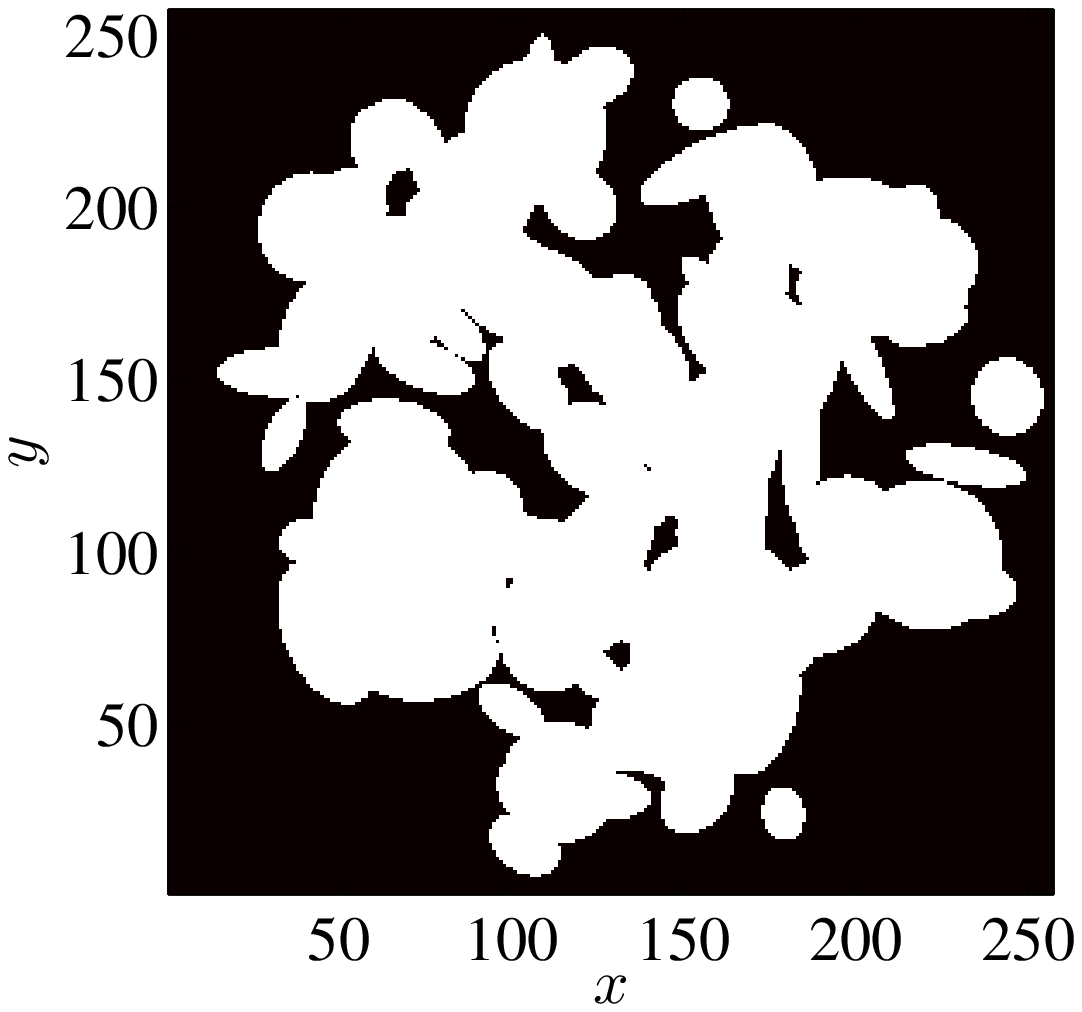}}
\\
\subfigure[$n=200$, $r_{min}=5$, $r_{max}=10$]
{\includegraphics[width=.9\columnwidth]{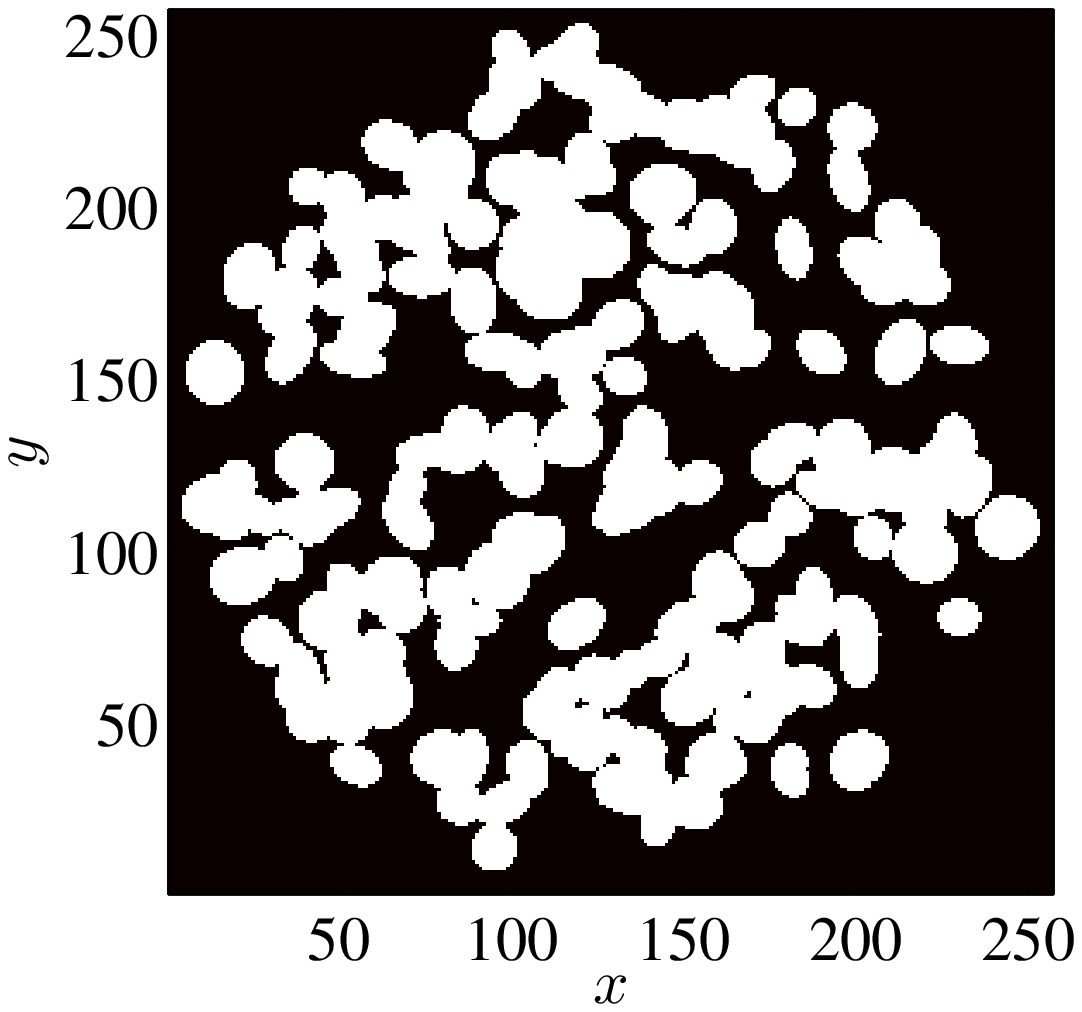}}
\end{center}
\caption{ An example for the ellipse test cases for the five sets of parameters
$(n,r_{min},r_{max})$. }\label{fig:TestEllips}
\end{figure*}

The second series of tests consists of the union of $n$ random
ellipses whose principal axes are picked at random in the interval
$[r_{min},r_{max}]$. The principal axis of the ellipse is also
chosen randomly with a uniform distribution over $[0,\pi]$ (\ie
isotropic distribution). Hence a test case is characterized by
three parameters $(n,r_{min},r_{max})$.
Figure~\ref{fig:TestEllips} shows an example of the five cases
considered in the following series.  Here again, the number $M$ of
projection directions was varied depending on the complexity of
the case. The image size was here again
$257~\times~257$~``pixels.''

As in the previous case, 200 samples were generated for each set
of parameters, and averages over these samples are reported in
Tables~\ref{Tab:Ellipse_Bat} (reproduced from
Ref.~\cite{Batenburg07}) and~\ref{Tab:Ellipse_Present} for the
proposed algorithm.  Let us note that we reproduce here only
global numbers and did not consider statistics over ``successful''
reconstructions.

\begin{table*}
\begin{center}
\caption{ Ellipse test case reproduced from
Ref.~\cite{Batenburg07}.}\label{Tab:Ellipse_Bat}
\begin{tabular}{ccclr@{.}lr@{.}lr@{.}lrr}
\hline
$n$ &$r_{min}$ &$r_{max}$ &$M$  &\multicolumn{2}{c}{\# perfect (\%)}&
\multicolumn{2}{c}{Proj. error} &\multicolumn{2}{c}{Pixel error} &Time(s)\\
\hline
15 &20 &40 &4   &~~~~38&5        &~~170&0 &~~3257&0    &36~~~~\\
&&         &5   &~~~~100&0          &~~0&0     &0&0    &19~~~~\\
&&         &6   &~~~~100&0          &~~0&0     &0&0    &13~~~~\\
\noalign{\smallskip\smallskip}
50 &5 &35  &5   &~~~~4&5          &~~737&0  &7597&0    &57~~~~\\
&&         &6   &~~~~56&5         &~~577&0  &2779&0    &43~~~~\\
&&         &7   &~~~~61&0           &~~5&9     &1&2    &22~~~~\\
&&         &8   &~~~~79&5           &~~7&5     &1&3    &20~~~~\\
\noalign{\smallskip\smallskip}
50 &5 &25  &6   &~~~~17&5        &~~1202&0  &6510&0    &54~~~~\\
&&         &7   &~~~~54&5         &~~521&0  &1289&0    &37~~~~\\
&&         &8   &~~~~81&0           &~~6&3     &1&1    &21~~~~\\
&&         &9   &~~~~65&0          &~~13&2     &1&9    &21~~~~\\
\noalign{\smallskip\smallskip}
100 &5 &25 &7   &~~~~5&5         &~~1589&0  &4455&0    &61~~~~\\
&&         &8   &~~~~31&5          &~~270&0  &457&0    &40~~~~\\
&&         &9   &~~~~36&5           &~~31&0    &4&5    &31~~~~\\
\noalign{\smallskip\smallskip}
200 &5 &10 &12  &~~~~2&5          &~~6699&0& 6157&0    &117~~~~\\
&&         &14  &~~~~13&5          &~~107&0    &9&0    &68~~~~\\
&&         &16  &~~~~13&0          &~~131&0    &9&5    &48~~~~\\
\hline
\end{tabular}
\end{center}
\end{table*}

\begin{table*}
\begin{center}
\caption{ Ellipse test cases with the proposed algorithm (number of levels $=3$,
$a_0=4$, $\alpha=0.87$).}\label{Tab:Ellipse_Present}
\begin{tabular}{ccclr@{.}lr@{.}lr@{.}lrr}
\hline
$n$ &$r_{min}$ &$r_{max}$ &$M$  &\multicolumn{2}{c}{\# perfect (\%)}&
\multicolumn{2}{c}{Proj. error} &\multicolumn{2}{c}{Pixel error} &Time(s)\\
\hline
15& 20& 40& 4&    83&5&                2&&    41&2&  2.0~~~~\\
&&&         5&    99&5&                 0&&    0&005&   1.4~~~~\\
&&&         6&~~~~100&0&                0&&      0&&   1.3~~~~\\
\noalign{\smallskip\smallskip}
50&  5& 35& 5&    73&0&             ~~19&&    497&&  6.3~~~~\\
&&&         6&    97&5&                2&&     15&&  5.1~~~~\\
&&&         7&    100&0&                 0&&    0&&  4.4~~~~\\
&&&         8&    99&5&                0&&      0&4&  4.4~~~~\\
\noalign{\smallskip\smallskip}
50&  5& 25& 6&    46&5&               43&&   1665&&  8.6~~~~\\
&&&         7&    97&0&                 2&&     45&&  5.6~~~~\\
&&&         8&    99&5&                 1&&     15&&  5.3~~~~\\
&&&         9&    100&0&                0&&      0&&  4.6~~~~\\
\noalign{\smallskip\smallskip}
100& 5& 25&  7&    90&5&                5&&     79&&  8.3~~~~\\
&&&         8&    99&0&                 1&&    10&&  7.9~~~~\\
&&&         9&    99&5&                 0&&    0&02&  8.1~~~~\\
\noalign{\smallskip\smallskip}
200& 5& 10&  12&   22&5&              152&&   2472&&  19.8~~~~\\
&&&         14&   98&5&                 3&&      5&&  14.1~~~~\\
&&&         16&   98&5&                 3&&      5&&  14.8~~~~\\
\hline
\end{tabular}
\end{center}
\end{table*}

The ellipse test cases are more demanding than the polygon ones,
and require a larger number of projections. The chosen number of
projections is clearly for the smaller values at the lower limit
of the minimum value for a decent reconstruction success rate.
Only the first series with fifteen ellipses and four projections
can be tackled safely with 100-\% success for 5 and 6 projections
in Table~\ref{Tab:Ellipse_Bat} from Ref.~\cite{Batenburg07}.  The
last series (200 ellipses) can hardly be considered as
satisfactory with at best 13-\% success rate. In contrast, we
observe that, with the present approach, all series reach a
success rate above 98.5-\% provided enough projections are taken
into account. The projection and pixel errors are systematically
much smaller with the present approach, even when the overall
success rate is low.  Finally the computation time is reduced by a
factor from 5 to 10 from that reported by
Batenburg~\cite{Batenburg07}.

For all the test cases, the proposed algorithm  is shown to
perform better than Batenburg's procedure both in terms of lower
residual error (projection and reconstructed image) and lower
computation time.

\section{Complexity} \label{Sec:Complexity}

Considering the previous results, let us try to formulate a
measurement of ``complexity'' relevant for our algorithm.  As
mentioned in the introduction, Donoho and
Tanner~\cite{Donoho10a,Donoho10b} have derived an appropriate
measurement of complexity in the case where no spatial correlation
are present in the image to be reconstructed.  In this case,
complexity has to be a function of the fraction of 1-pixels, $p$,
(or 0-pixels whichever is the minority value). Moreover, a
critical value $p_c$ given in Eq.~\ref{eq:pc_Donoho} allows one to
distinguish solvable problems where $p<p_c$ from those where
information is lacking, $p>p_c$.  \revisionter{Thus $\chi_A=p/p_c=
2p \delta\log(\delta)$ can be seen as a measure of complexity for
Type A problems where $\delta=N/M$ is the undersampling ratio.}

\begin{figure}[ht!]
\begin{center}
{\includegraphics[width=.9\columnwidth]{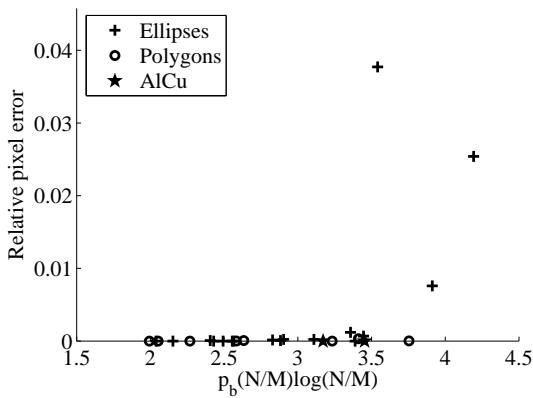}}
\end{center}
\caption{Relative pixel error as a function of the
proposed measure of complexity, $\chi_B$.}\label{fig:Complexity}
\end{figure}

No such result has been derived for Type B problems, where
boundaries play somehow the role played by minority pixels in Type
A.  If the fraction of boundary ``bonds'' (\ie pairs of
neighbouring pixels having a different value) $p_b$ is introduced,
\revisionter{it is natural to consider
    \be
    \chi_B=p_b\delta\log(\delta)
    \ee
as a candidate for quantifying the image complexity for Type B.}
Figure~\ref{fig:Complexity} is an attempt to quantify the final
pixel error as a function of the average $\chi_B$ measured from
each image series.  On this graph, all the test cases reported
earlier (polygons and ellipses) have been collected, as well as
the AlCu tomographic image.  This plot seems to indicate that
small values of $\chi_B\le3.5$ can be solved exactly, while values
above 4 may be out of reach of the present algorithm.  Note
however that there is no formal justification for this choice
concerning the measure of complexity.

\section{Robustness with respect to noise}
\label{Sec:Noise}

To test the robustness of the proposed methodology with respect to
noise, only the case of the microstructure shown in
Figure~\ref{fig:AlCu} has been chosen as representative of a
challenging case, in particular considering its size (\ie
1024$\times$1024 pixels), and its complex geometry.  The physics
of noise generation in the projections reflects a number of
different phenomena. The ambition is not to reproduce a specific
noise, but rather to test the robustness of the algorithm.  Thus
it is chosen to add a Gaussian white noise to the projection data.
The noise is characterized by its Signal to Noise Ratio, SNR, such
that
    \be
    \mathrm{SNR}=20\log_{10}\left(\frac{\langle \pi \rangle}{\eta}\right)
    \ee
where $\eta$ is the standard deviation of the noise.

The number of projections is $M=15$, and a standard set of
parameters was chosen to deal for all SNRs.  The number of scales
was chosen to be 4, with 30 iterations per scale. The
regularization length scale was chosen to be larger than in the
preceding examples $a_0=10$, and hence with a faster decay rate
$\alpha=0.8$. At the end of the prescribed number of iterations,
the relative pixel and projection errors are computed. Relative
means here that the pixel error is normalized by the total number
of pixels in the reconstructed disk, and the projection error is
normalized by the mean projection data, $\langle \pi \rangle$.

\begin{figure}[ht!]
\begin{center}
{\includegraphics[width=.9\columnwidth]{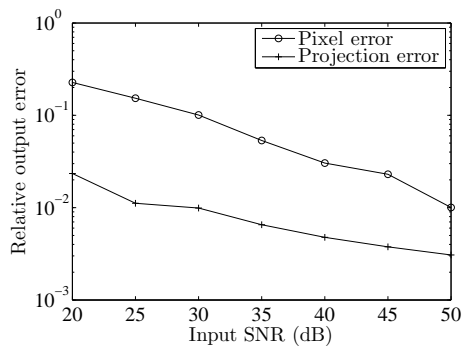}}
\end{center}
\caption{Evolution of pixel or projection errors at
the end of the reconstruction as a function of the noise signal
ratio, SNR.}\label{fig:Noise}
\end{figure}

Figure~\ref{fig:Noise} shows the change of both errors as
functions of the SNR.  It is observed that both errors have a
similar evolution, and although the process is intrinsically
susceptible to noise (because of the low level of information
provided), convergence to decent reconstruction is still achieved.
For a 1\% level of fluctuations in the projection data (SNR$=40$),
the relative pixel error is about 3\%. Let us emphasize the fact
that spatial regularization is essentially active in the first few
iterations (when $a$ is significantly larger than 1) and hence no
filter is applied in the last stages apart from the projection
constraints that are corrupted by noise.  A post-processing of the
data to remove isolated pixels for instance (or involving a more
elaborate filtering) is able to reduce significantly the resulting
error. This section is included to document the robustness of the
proposed procedure, but the latter has not been designed to be
especially immune to noise, but rather to allow for an efficient
reconstruction for a low number of projections $M$.


\section{Conclusion}\label{Sec:Conclusion}

A novel binary reconstruction algorithm has been introduced.  It
is based on a nonlinear transformation of the probability for a
site to be 1-valued.  The proposed initialization step is a mere
backprojection applied to a non-linear transformed projection
data, followed by a correction step. A correction scheme to meet
the projection constraints and involving a minimal regularization
procedure (binarization and convolution by a small width Gaussian
kernel) allows error-free binary reconstructions to be achieved
for a variety of examples of large image sizes (examples of
1-Mpixel images have been shown). A multiscale version speeds up
convergence. A detailed comparison with a series of test cases
provided in Ref.~\cite{Batenburg07} shows that the proposed
algorithm achieves very good performances (superior success rate
and lower computation time).

Let us emphasize that the number of required projection data can
be as low as 1~\% of the required number for standard
reconstructions.  This has direct consequences in terms of X-ray
dose reduction for medical applications (provided the severe
constraint (or approximation) or looking for a binary image can be
acceptable), or for fast-imaging where acquisition time is
limited.

An appealing route is to provide an efficient computer
implementation of the proposed algorithm.  Up to now a basic
Matlab$^{\circledR}$ code has been used on a simple processor PC.
GPU implementation is expected to offer much higher performances.

Finally, extensions of such approaches to discrete (rather than
binary) images is also very challenging as it would open new
perspectives for applications, going beyond the very restrictive
frame of binary images.  In a similar spirit, the application of
the proposed binary algorithm to images that are not binary is an
interesting question to be addressed.

\begin{acknowledgements}

Communication of the raw tomographic data shown in
Figure~\ref{fig:AlCu} by E.~Gouillart (CNRS/Saint-Gobain,
Aubervilliers, France) and C.~Zang (RWTH, Aachen, Germany) is
gratefully acknowledged. We are also indebted to E.~Gouillart for
the suggestion that boundary sites may be an appropriate measure
of complexity as proposed in Ref.~\cite{Gouillart}.
\revisionter{We also thank an anonymous reviewer for interesting
and constructive remarks.} This work is supported by the French
Agence Nationale de la Recherche through ``RUPXCUBE''
(ANR-09-BLAN-0009-01) and ``EDDAM'' (ANR-11-BS09-027) projects.

\end{acknowledgements}

\end{document}